\documentclass[12pt]{iopart}

\pdfoutput=1

\usepackage[english]{babel}
\usepackage{
amssymb,amsfonts,latexsym}
\usepackage{graphicx}
\usepackage{color}
\usepackage{iopams}
\usepackage{subfigure}
\usepackage{afterpage}
\usepackage{booktabs}
\usepackage{blkarray}
\usepackage{cite}
\usepackage{tikz}
\usetikzlibrary{snakes}
\usepackage{soul}
\usetikzlibrary{calc}
\usepackage{bm}
\usepackage{braket}

\definecolor{blue(pigment)}{rgb}{0.2, 0.2, 0.6}
\usepackage{mathrsfs}
\usepackage{times}
\usepackage[colorlinks,
citecolor=blue,linkcolor=blue,urlcolor=blue]{hyperref}
\hypersetup{linktocpage}
\usepackage{scalerel}
\usepackage{tensor}
\usepackage{etoolbox}

\makeatletter
\def\@mkboth#1#2{}
\newlength\appendixwidth
\preto\appendix{\addtocontents{toc}{\protect\patchl@section}}
\newcommand{\patchl@section}{%
  \settowidth{\appendixwidth}{\textbf{Appendix }}%
  \addtolength{\appendixwidth}{1.5em}%
  \patchcmd{\l@section}{1.5em}{\appendixwidth}{}{\ddt}%
}
\makeatother

\def\eqref#1{(\ref{#1})}
\usepackage{cancel}

\newcommand{\LL}{\zeta}
\definecolor{green4}{rgb}{0.0, 0.5, 0.0}

\newcommand{\bea}{\begin{eqnarray}}
\newcommand{\eea}{\end{eqnarray}}

\newcommand\reallywidehat[1]{\arraycolsep=0pt\relax%
\begin{array}{c}
\stretchto{
  \scaleto{
    \scalerel*[\widthof{\ensuremath{#1}}]{\kern-.5pt\bigwedge\kern-.5pt}
    {\rule[-\textheight/2]{1ex}{\textheight}} 
  }{\textheight} %
}{0.5ex}\\           
#1\\                 
\rule{-1ex}{0ex}
\end{array}
}
\newcommand{\A}{\hat{\cal A}}
\newcommand{\reg}{\mathrm{\small{reg}}}

\newcommand{\x}{\mathrm{x}}
\newcommand{\cp}{c^\prime}
\newcommand{\Psik}{\hat\Psi_\kappa}
\newcommand{\Psikd}{\hat\Psi_\kappa^\dagger}
\newcommand{\sgn}{\mathrm{sgn}}
\newcommand{\limTh}{\mathrm{limTh}}
\newcommand{\gammap}{\gamma^\prime}
\newcommand{\xp}{x^\prime}

\newcommand*{\ddt}[1]{%
  \accentset{\mbox{\bfseries .\hspace{-0.25ex}.}}{#1}} 

\newcommand{\To}{\rightarrow}
\newcommand{\be}{\begin{equation}}
\newcommand{\ee}{\end{equation}}

\newcommand{\beA}{\begin{align}}
\newcommand{\eeA}{\end{align}}

\newcommand{\bV}{\begin{bmatrix}}
\newcommand{\eV}{\end{bmatrix}}

\newcommand{\dd}{\mathrm{d}}
\newcommand{\de}{\partial}

\begin{document}

\title[]{One-particle density matrix of a trapped Lieb-Liniger anyonic gas}
\author{Stefano Scopa$^{1}$, Lorenzo Piroli$^{2,3}$, Pasquale Calabrese$^{1,4}$}
\address{$^1$ SISSA and INFN, via Bonomea 265, 34136 Trieste, Italy}
\address{$^2$ Max-Planck-Institut f\"ur Quantenoptik, Hans-Kopfermann-Str. 1, 85748 Garching, Germany}
\address{$^3$ Munich Center for Quantum Science and Technology, Schellingstraße 4, 80799 M\"unchen, Germany}
\address{$^4$ International  Centre  for  Theoretical  Physics  (ICTP),  I-34151,  Trieste,  Italy}
\date{\today}
\begin{abstract}
We provide a thorough characterisation of the zero-temperature one-particle density matrix of trapped interacting anyonic gases in one dimension, 
exploiting recent advances in the field theory description of spatially inhomogeneous quantum systems. 
We first revisit homogeneous anyonic gases with point-wise interactions. 
In the harmonic Luttinger liquid expansion of the one-particle density matrix for finite interaction strength, the non-universal field 
amplitudes were not yet known. We extract them from the Bethe Ansatz formula for the field form factors, providing an exact asymptotic 
expansion of this correlation function, thus extending the available results in the Tonks-Girardeau limit. 
Next, we analyse trapped gases with non-trivial density profiles. 
By applying recent analytic and numerical techniques for inhomogeneous Luttinger liquids,
we provide exact expansions for the one-particle density matrix. We present our results for different confining potentials, 
highlighting the main differences with respect to bosonic gases.

\end{abstract}

\maketitle

\tableofcontents

\section{Introduction}
\label{sec:intro}

The concept of indistinguishable particles is one of the defining features of quantum mechanics. 
{An old and fundamental result is} that in three spatial dimensions indistinguishability is only compatible with bosonic and fermionic statistics for 
elementary particles. 
Conversely, two dimensions also bear anyons which are particles with  features interpolating between the two standard statistics~\cite{lm_77,laughlin_83}. 

{A number of recent} works have suggested the possibility of observing anyons also in one spatial dimension, within carefully engineered cold atomic setups ~\cite{klmr-11,gs-15,sse-16,gcs-18}. These works complemented previous theoretical investigations, where different models of 1$D$ anyonic particles were introduced ~\cite{agjp-96,rabello-96,it-99,Kundu99,lmp-00,Girardeau06,Batchelor06,an-07,Patu07,Batchelor2007,BeCM09,yao-12,spk-12,bfglz-08,zgf-17,yp-17,jy-18,rccm-19,mwd-20},
exhibiting intriguing features that {are not present} in fermionic or bosonic systems.

In practice, 1$D$ anyons can be described by quantum field operators $\Psik(x)$, $\Psik^\dagger(x)$ satisfying the generalised commutation relations
\bea\label{anyon-rules1}
&\Psik^\dagger(x_1)\Psikd(x_2)= e^{i\pi\kappa\,\sgn(x_1-x_2)}\; \Psikd(x_2)\Psikd(x_1)\,,\label{eq:comm_1}\\[3pt]\label{anyon-rules2}
&\Psik(x_1)\Psikd(x_2)=e^{-i\pi\kappa\,\sgn(x_1-x_2)}\; \Psikd(x_2)\Psik(x_1)+ \delta(x_1-x_2)\,,\label{eq:comm_2}
\eea 
where ${\rm sgn}(x)$ is the sign function, while $\kappa\in [0,1]$ is the statistical parameter, {with $\kappa=0$ corresponding to bosons and $\kappa=1$ to fermions. 
Most of the works in the literature concern non-interacting anyons, but interactions also lead to interesting phenomena. 
In one of the simplest instances}, the interaction between anyons is described by a delta-like potential, leading to the anyonic Lieb-Liniger gas~\cite{Patu07,Batchelor2007}, which {is also the leading character of} this work.
 {Another famous model of interacting anyonic particles is the Calogero-Sutherland model \cite{CSM1,CSM2}, where, however, a fractional statistic emerges from collective excitations rather than from the requirement of generalized commutation relations for the fields (cf. Eq.~\eqref{anyon-rules1}-\eqref{anyon-rules2}).}

As its bosonic counterpart~\cite{Lieb63}, the anyonic Lieb-Liniger model is integrable \cite{Kundu99}. Despite its apparent simplicity, it displays several interesting features, and a series of works have already achieved a precise characterisation of its spectral and thermodynamic properties~\cite{Patu07,Batchelor2007}. On the other hand, the computation of  correlation functions turned out to be much more challenging, as it is notoriously the case for Bethe-Ansatz solvable models~\cite{Korepin1993}. 
For instance, analytic results for the bosonic model could be obtained only after many years of technical advances, leading to exact formulas at zero~\cite{JiMi81,OlDu03,GaSh03,CaCa06,ChSZ05,ChSZ06,ScFl07,PiCa16-1,km-15,kkmt-13,kt-11} and finite temperatures~\cite{MiVT02,KGDS03,SGDV08,KuLa13,PaKl13,ViMi13,PaCa14,NRTG16,kms-11,kms-11b}, and, more recently, for arbitrary excited states~\cite{KoMT09,KoCI11,KoMT11_sTG,Pozs11_mv,Pozs11,PiCE16,BaPi18,BaPC18,CaCaSl07,KoMP10,PiCa15}. In the anyonic model, a generalisation of most of these results is still lacking, and the vast majority of  existing studies is restricted to the limit of infinitely repulsive interactions (aka Tonks-Girardeau gas), 
which can be analysed by means of an anyon-fermion mapping~\cite{Girardeau06}. In this regime, quantitative predictions for correlation functions and (particle) {entanglement} entropy have been obtained both in~ \cite{Calabrese2007,Patu08,Patu09,Calabrese2009,Patu19,Santachiara2007,ghc-09,Santachiara2008,hzc-08,Patu09b,Patu15,Zinn15,hao-16,Marmorini16,hs-17,tep-15} 
and out  of equilibrium~\cite{del_Campo08,hc-12,Li-15,wrdk-14,PiCa17,fgd-18}. 

{A part the per se interest,} the computation of correlation functions {for finite interaction strength is also needed}
in view of possible experimental implementations.  
Furthermore, cold atomic setups necessarily require to take into account confining potentials~\cite{BlDZ08,GuBL13}, which break the integrability of the model. 
{Consequently, in these inhomogeneous situations,} Bethe Ansatz techniques alone are not powerful enough to provide quantitative predictions.

{In thermal equilibrium at low energy (in particular at zero temperature)}, these difficulties can be partially overcome by exploiting a Luttinger liquid description~\cite{Haldane1981a,Haldane1981b,Haldane1981c,Gogolin-Tsvelik,Giamarchi2003,Cazalilla2004}. The latter  represents a well-known hydrodynamic approach, which {provides} the universal long-range behaviour of correlation functions. 
A key point is that, while Luttinger liquids (and, more generally, conformal field theory descriptions) have been traditionally employed in homogeneous situations, over the past few years a series of studies have established the possibility of extending their application to inhomogeneous settings~\cite{MaSt95,Kats11,HiNi11, WeRL16,RDRC17,DuSC17,ToRS18,ADSV16,Dubail17,Eisler17,Brun2017,Brun2018,Bastianello2020,Y-thesis,j-lec,tss}. 
{All these advances place at our disposal a set of versatile tools} to study interacting systems in the presence of trapping potentials, bridging further the gap between theory and experiments.

{Luttinger liquid techniques have been already exploited to describe the correlation functions of a uniform anyonic Lieb-Liniger gas for finite interaction strength 
\cite{Calabrese2007}.
However, this investigation left open the determination of the non-universal field amplitudes which are always out of reach of universal conformal approaches}. Indeed,  in this framework correlation functions  are expressed in terms of non-universal parameters, that should be fixed {from independent microscopic} calculations. In  the bosonic Lieb-Liniger model, this {computation was performed exploiting exact Bethe Ansatz formulas for the matrix elements (or \emph{form factors}) of local operators~\cite{SGCI11,SPCI12}. However, in the anyonic case, a generalisation of some of these formulas only appeared this year in Ref.~\cite{Piroli2020}.

The aim of this work is to put together the exact results of Ref.~\cite{Piroli2020}, and the formalism of inhomogeneous Luttinger Liquids developed in Refs.~\cite{Brun2017,Brun2018,Bastianello2020,Y-thesis,j-lec} to present a study of the one-particle density matrix in the anyonic Lieb-Liniger gas. 
{En route}, we extract the non-universal coefficients appearing in the Luttinger liquid description based on the {form factors} of Ref.~\cite{Piroli2020}. 
We  present a series of exact calculations both for a homogeneous gas and in the presence of confining potentials. 
To our knowledge, this work provides the first predictions (without the need of any fitting parameters) for the  
correlation functions of anyonic $1D$ gases beyond the Tonks-Girardeau limit.

The rest of this manuscript is organised as follows. We start in Sec.~\ref{sec:model} by introducing the anyonic Lieb-Liniger model and its Bethe Ansatz solution, while the Luttinger liquid approach is reviewed in Sec.~\ref{sec:luttinger-liquid}. In Sec.~\ref{sec:corr-homo} we present our quantitative results for the one-particle density matrix in the homogeneous case, while sections~\ref{sec:conf_pot} and \ref{sec:opf_in_setup} are devoted to the analysis of the same quantity in the presence of confining potentials. Finally, our conclusions are reported in Sec.~\ref{sec:conclusions}.  The most technical aspects of our work are consigned to several appendices.

\section{The anyonic Lieb-Liniger model}
\label{sec:model}

We start by introducing the anyonic Lieb-Liniger model~\cite{Kundu99,Batchelor06,Batchelor2007}, describing a gas of anyonic particles with point-wise repulsive interactions, and confined on a one-dimensional ring of length $L$. The Hamiltonian reads
\be\label{anyon-Lieb-Liniger}
\hat{H}=\int_{0}^{L} \dd x \left[ \de_x \hat\Psi^\dagger_{\kappa}(x)\ \de_x\ \hat\Psi_{\kappa}(x) +c \,\hat\Psi_\kappa^\dagger(x)^2\ \hat\Psi_\kappa(x) ^2\right]\,.
\ee
The anyonic fields  $\Psik^\dagger$, $\Psik$ satisfy the generalised commutation relations introduced in Eqs.~\eqref{eq:comm_1} and \eqref{eq:comm_2}. 
{We recall} that the anyonic parameter $\kappa$ is equal to $0$ for bosons and $1$ for spinless fermions.  
The Hamiltonian \eqref{anyon-Lieb-Liniger} generalises to anyons the well-known bosonic Lieb-Liniger model \cite{Lieb63}. It was introduced and solved using the Bethe Ansatz by Kundu \cite{Kundu99}, and systematically analysed by Batchelor {\it et al.} \cite{Batchelor06,Batchelor2007} and P\^{a}tu {\it et al} \cite{Patu07,Patu08,Patu09}. 
In the following, we briefly review  the main features of its exact solution.
\subsection{Bethe Ansatz solution}\label{sec:Bethe-Ansatz}
We denote by $\ket{\chi_N}$ a $N$-particle state of the form
\be
\fl \ket{\chi_N}=\frac{1}{\sqrt{N !}} \int_{0}^{L} \dd x_1 \dots \int_{0}^{L} \dd x_N \; \chi_N(x_1,\dots,x_N)\, \Psikd(x_1) \,\dots\, \Psikd(x_N) \, \ket{0}\,,
\ee 
where $\ket{0}$ is the Fock vacuum and $\chi_N$ is the many-body wavefunction satisfying
\be
\chi_N(\dots,x_j,x_{j+1},\dots)= e^{i\pi\kappa \sgn(x_j-x_{j+1})}\, \chi_N(\dots,x_{j+1},x_j,\dots),
\ee
under particle exchange. In the following, we impose periodic boundary conditions for the anyonic field, namely $\Psikd(L)=\Psikd(0)$. As discussed in Ref.~\cite{Patu07}, due to the anyonic commutation relations, this {request} does not imply periodicity of the wavefunction in all the coordinates $x_j$. Instead, a consistent choice, which will be employed in this work, is~\cite{Patu07}
\bea\label{boundary}
&\chi_N(0,x_2,\dots,x_N)= \chi_N(L,x_2,\dots,x_N)\,,\nonumber\\[3pt]
&\chi_N(x_1,0,\dots,x_N) = e^{i2\pi\kappa} \chi_N(x_1,L,\dots,x_N)\,,\nonumber\\[3pt]
&\qquad \vdots\nonumber\\[3pt]
&\chi_N(x_1,x_2,\dots,0)= e^{i2\pi\kappa (N-1)}\, \chi_N(x_1,x_2,\dots,L)\,.
\eea

The eigenvalue problem $\hat{H}\ket{\chi_N}=E \ket{\chi_N}$ can be rewritten in the language of first quantisation as
\be
\left(-\sum_{j=1}^N \frac{\de^2}{\de x_j^2} + 2c \sum_{1\leq j\leq k \leq N} \delta(x_j-x_k)\right) \chi_N= E_N \; \chi_N,
\ee 
with boundary conditions given by Eq.~\eqref{boundary}. A complete solution was obtained in Refs.~\cite{Kundu99,Batchelor2007,Patu07} by a standard application of the Bethe Ansatz. In particular, the wavefunction of a given $N$-particle eigenstate {is}
\bea\label{eigenstate}
\chi_N(x_1,\dots,x_N)=&\frac{e^{i\frac{\pi\kappa}{2}\sum_{j<k} \sgn(x_j-x_k)}}{\sqrt{N ! \prod_{j>k}\left[(\lambda_j-\lambda_k)^2+c^{\prime \ 2}\right]}}\, \sum_{{\cal P}\in S_N} (-1)^{\cal P} e^{i\sum_{j=1}^N x_j \lambda_{{\cal P}_j}}\nonumber\\
&\times\prod_{j>k} \left[\lambda_{{\cal P}_j} -\lambda_{{\cal P}_k} - i\cp \,\sgn(x_j-x_k)\right]\,,
\eea
where ${\cal P}$ denotes a permutation of $N$ indices, while 
\be\label{cp}
\cp\equiv \frac{c}{\cos(\pi \kappa/2)},
\ee
is the effective coupling. Here $\{\lambda_j\}_{j=1}^N$ are a set of  of quasimomenta (or \emph{rapidities}) parametrising the different eigenstates, and satisfying the Bethe equations
\be\label{BAE}
e^{i\lambda_j L}= e^{-i\pi \kappa (N-1)} \prod_{k\neq j, k=1}^N \left(\frac{\lambda_j-\lambda_k +i c^\prime}{\lambda_j-\lambda_k-i c^\prime}\right)\,,
\ee
which can be conveniently rewritten in logarithmic form as
\be\label{log-BAE}
\lambda_j \, L = 2 \pi I_j -2\pi\{\pi\kappa (N-1)\}_{2\pi} -2 \sum_{k=1}^N \arctan\left(\frac{\lambda_j-\lambda_k}{\cp}\right)\,.
\ee
 Here we introduced the quantum numbers $I_j$, which must be chosen to be pair-wise distinct and integers (semi-integers) for $N$ even (odd). The ground state corresponds to
\be
 \{I_j\}_{j=1}^N=\left\{-\frac{(N-1)}{2},-\frac{(N-1)}{2}+1,\ldots, \frac{N-1}{2} \right\}\,.
 \label{eq:qn_gs}
 \ee
Following \cite{Patu07}, we also introduced the notation
\be
\{ x\}_{2\pi}\equiv s  \quad{\rm if} \quad x =2\pi m + 2\pi s, \quad  s\in[0,1)
\ee
with $m$ an integer.\\
Finally, given a solution of the Bethe equation \eqref{BAE}, the total energy and momentum of the eigenstate read
\be
E[\{\lambda_j\}_{j=1}^N]=\sum_{j=1}^N \lambda_j^2, \qquad P[\{\lambda_j\}_{j=1}^N]=\sum_{j=1}^N \lambda_j.
\label{EPBA}
\ee
\indent

The Bethe Ansatz solution outlined above allows for a straightforward definition of the thermodynamic limit  ($\limTh$), 
in analogy with the well-known bosonic case~\cite{Korepin1993}.
Eq.~\eqref{log-BAE} can be written in an integral form, when $N, L \To \infty$ at fixed density $\rho=N/L$. 
In particular {for the ground state, with quantum numbers} in Eq.~\eqref{eq:qn_gs}, we obtain the  Lieb equation \cite{Batchelor2007}
\be\label{TBA}
2\pi\rho_p(\lambda)= 1+ \int_{-Q}^{Q} \dd\mu \, \frac{2\cp}{c^{\prime\ 2} + (\lambda-\mu)^2}\, \rho_p(\mu)\,.
\ee
Here $\rho_p(\lambda)$ is the rapidity distribution function (or {\it``root density''}), which generalises the concept of momentum occupation number to the interacting case. 
Formally, it is defined as $\rho_p(\lambda)=\limTh\, (L(\lambda_{j+1}-\lambda_j))^{-1}$. 
The shift due to $\kappa$ in the rapidities solving Eq.~\eqref{log-BAE} at finite size vanishes as $1/L$ in the thermodynamic limit \cite{Batchelor06}. 
The extreme $\pm Q$ of the integration are self-consistently obtained from the equation for the particle density
\be
\rho=\int_{-Q}^{Q} \dd \lambda \ \rho_p(\lambda)\,. 
\ee
 The total energy is instead 
\be
e=\int_{-Q}^{Q} \dd \lambda \ \rho_p(\lambda)\lambda^2\,.
\ee
For later use we also introduce the pseudoenergy function $\epsilon(\lambda)$ satisfying the integral equation
\be\label{pseudo}
\epsilon(\lambda)=\lambda^2 +\frac{1}{2\pi} \int_{-Q}^{Q} \dd\mu\,  \frac{2\cp}{c^{\prime\ 2} + (\lambda-\mu)^2}\, \epsilon(\mu),
\ee
which is useful to fix $Q$ when working at finite chemical potential by requiring that $\epsilon(\pm Q) =0$.

\section{Luttinger liquid approach for anyons}
\label{sec:luttinger-liquid}

While the Bethe Ansatz provides a simple characterisation of the spectrum of the Hamiltonian and of the thermodynamics of a model, the computation of correlation functions turns out to be a much harder task. If, for example, one would use directly the ground state eigenfunction~\eqref{eigenstate} to generate a correlation, the complexity of this computation would grow exponentially with $N$, limiting the calculation to a handful of particles. Also the exact sum over the form factors of a given operator is growing exponentially with $N$, although one can arrange the intermediate states in order of relevance \cite{CaCaSl07}. 
The Luttinger Liquid \cite{Haldane1981a,Haldane1981b,Haldane1981c,Giamarchi2003,Cazalilla2004,Gogolin-Tsvelik} 
represents a viable field theory approach to overcome these difficulties at low energy and in the limit of large spatial distances, as we review in this section.

In one-dimension, the anyonic field $\Psikd$ can be represented in terms of a bosonic one $\hat\Psi_0^\dagger$ with {the help of} the transformation \cite{Calabrese2007}
\be\label{JW}
\Psikd(x)=\hat\Psi_0^\dagger(x) \, \exp\left(i\pi\kappa\int_{0}^x \dd y \, \hat\rho(y)\right),
\ee
where $\hat\rho=\Psikd\Psik=\hat\Psi^\dagger_0\hat\Psi_0$ is the particle density operator and $\exp(i\pi\kappa\int \dd x \hat\rho(x))$ is the statistical phase that characterises particles with anyonic statistics. Note that $\kappa=1$ reproduces  the Jordan-Wigner transformation between bosons and spinless fermions.
Then, we consider a coarse-graining procedure by writing the density operator as
\be
\hat\rho(x)=\hat\rho_>(x)+\hat\rho_<(x),
\ee
where $\hat\rho_{<}$ ($\hat\rho_>$) refers to long (short) wavelength modes occurring over distances $|x|\gg \rho^{-1}$ ($|x|\ll \rho^{-1}$), with $\rho=\braket{\hat\rho}$. 
At low energy, we can integrate out the fast modes $\hat\rho_>$ and retain only the long wavelength part $\hat\rho_<$. 

On physical grounds, fast modes $\hat\rho_>$ are expected to cancel inside the integral in Eq.~\eqref{JW}. 
Therefore, we can replace $\hat\rho$ with $\hat\rho_<$ in Eq.~\eqref{JW} obtaining 
\be\label{JW2}
\Psikd(x) = \hat\Psi_0^\dagger(x) \, e^{i\kappa \hat\Phi(x)},
\ee
in terms of the field $\hat\Phi$, which satisfies
\be\label{slow-modes}
\frac{1}{\pi} \de_x\hat\Phi(x)= \hat\rho_<(x).
\ee
One may interpret $\hat\Phi$ as a {field that keeps track of the positions of the finite number of particles in a one-dimensional configuration. Explicitly,}   
 the field $\hat\Phi$ {can be thought as a piecewise constant function of the position that  jumps of $\pi$ everywhere there is a particle, 
starting from $x=0$ up to  $x=L$}, see Fig.~\ref{fig:height-field} {for an illustration. 
In terms of the field $\hat\Phi$,} the density operator $\hat\rho(x)$ is written as the harmonic expansion \cite{Haldane1981a,Cazalilla2004,Giamarchi2003}
\be\label{density-exp}
\hat\rho(x)  \simeq  \de_x\hat\Phi \sum_{m=-\infty}^\infty \delta(\hat\Phi(x)-m\pi)  = \hat\rho_<(x) \sum_{m=-\infty}^\infty e^{2im \hat\Phi(x)}.
\ee
\begin{figure}[t]
\centering
\includegraphics[width=0.7\textwidth]{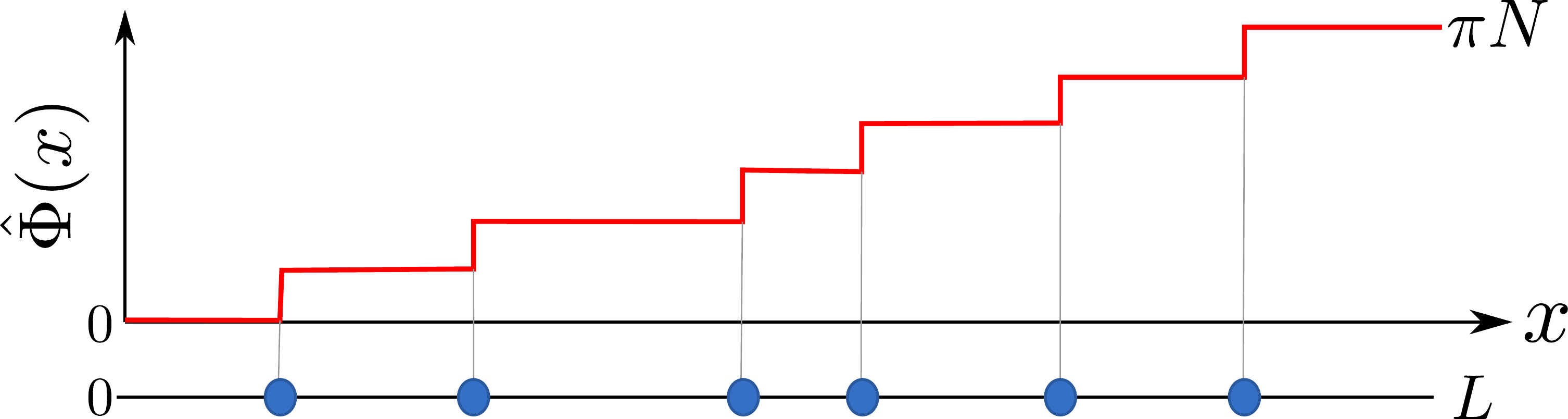}
\caption{Illustration of the field $\hat\Phi$ for a given configuration of particles.}\label{fig:height-field}
\end{figure}

We can now use the standard density-phase representation of the bosonic field $\hat\Psi^\dagger_0(x)\propto[\hat\rho(x)]^{1/2} e^{-i\hat\theta(x)}$.
The phase $\hat\theta(x)$ is the field conjugate to the density fluctuation 
\be
\hat\phi(x)\equiv\hat\Phi(x)-\pi\rho x,
\ee
with commutation rule $[\hat\theta(x), \hat\phi(x')]= i\pi \sgn(x - x')/2$. 
Plugging these definitions in the anyonic field~\eqref{JW2}, we can write $\Psikd(x)$ as 
\be\label{bosonization}
\Psikd(x) = \sqrt{\rho}\sum_{m=-\infty}^{\infty} B^{\kappa}_m  \, e^{i(2m+\kappa)\pi\rho x}\,  e^{i(2m+\kappa)\hat\phi(x)} \ e^{-i\hat\theta(x)},
\ee
with $\hat\rho_<(x)\simeq \rho$ and where we introduced the {constants} $B^{\kappa}_m$ as non-universal amplitudes.

\begin{figure}[t]
\centering
\includegraphics[width=0.7\textwidth]{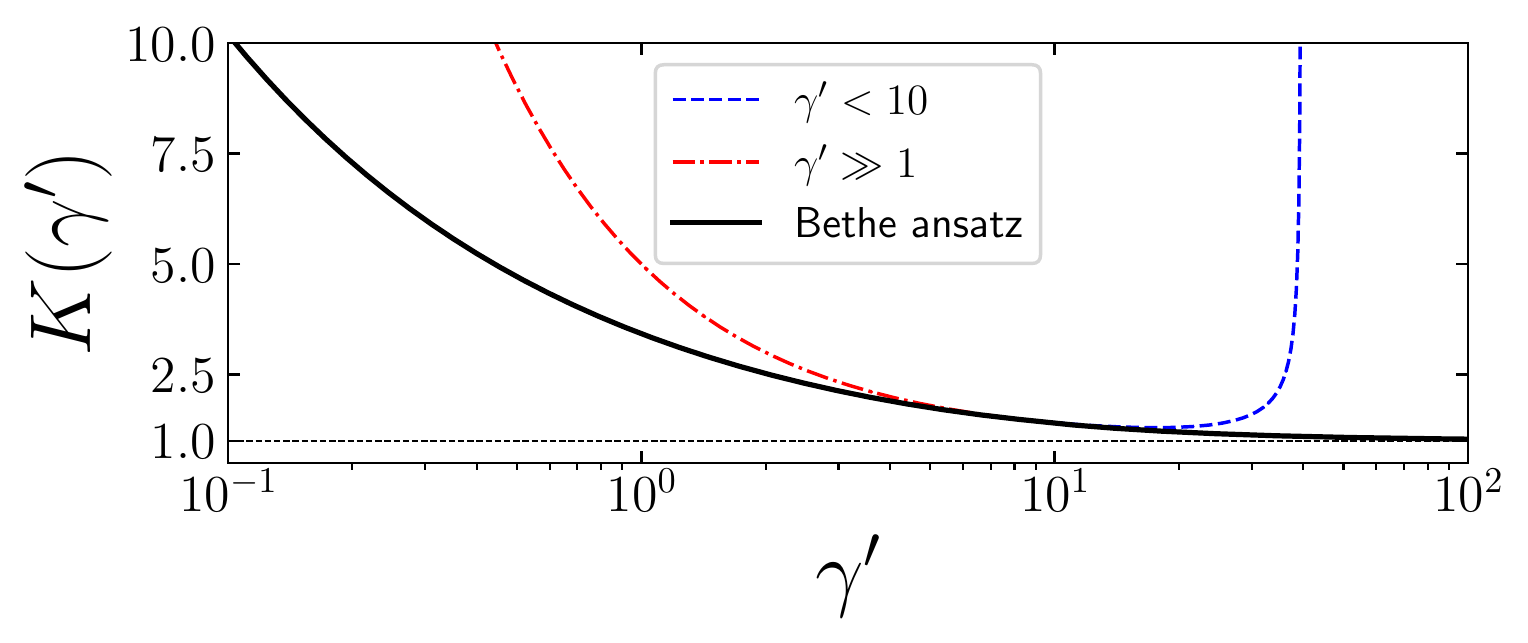}
\caption{The Luttinger parameter $K(\gamma^\prime)$ for the anyonic Lieb-Liniger model \eqref{anyon-Lieb-Liniger} as a function of $\gamma^\prime=\cp/\rho$ obtained from the numerical solution of Bethe Ansatz equations \eqref{log-BAE} and the evaluation of \eqref{vN} ({\it full line}). 
{The asymptotic behaviour for small $\gamma^\prime\lesssim 10$ is well described by the analytical result  
$K\simeq \sqrt{\pi/\gamma^\prime}(1-\sqrt{\gamma^\prime}/2\pi)^{1/2}$ \cite{Haldane1981a,Cazalilla2004} ({\it dashed line}).
For $\gamma^\prime \gg 1$ the asymptotic expansion is $K\simeq (1+ 4/\gamma^\prime)$} \cite{Cazalilla2004,Batchelor2007} ({\it dot-dashed line}).
}\label{fig:K}
\end{figure}

{The essence of the Luttinger liquid is that in terms of $\hat\phi$ and $\hat\theta$ the Hamiltonian is quadratic} \cite{Haldane1981a,Giamarchi2003}
\be\label{luttinger}
\hat{H}_{LL}=\frac{v_s}{2\pi} \int_{0}^{L} \dd x\,  \left(K\,(\de_x\hat\theta)^2 +\frac{1}{K}(\de_x\hat\phi)^2\right).
\ee
Here, $v_s$ is the sound velocity and $K$ is {known as} the Luttinger parameter. 
They are the only two parameters determining the low energy physics of the quantum fluid \cite{Haldane1981a}. 
Their actual value should be fixed from the exact solution of the microscopic model. 
In our case, they are just functions of the rescaled interaction coupling $\gammap=\cp/\rho$ and so readable from the bosonic result.
They are usually parametrised as 
\be
K=\sqrt{v_F/v_N}, \qquad {\rm and} \qquad v_s=\sqrt{v_F  v_N}=v_F/K,
\ee 
where $v_F=2\pi\rho$ is the Fermi velocity of spinless fermions and  $v_N$ is the density-stiffness of the gas \cite{Haldane1981a,Cazalilla2004}
\be\label{vN}
v_N=\frac{L}{\pi}\;\frac{\de^2 E_{\rm GS}(N)}{\de N^2}\Big\vert_{N=\rho L}.
\ee
The ground-state total energy $E_{\rm GS}(N)$ is defined as $E_{\rm GS}(N)=E[\{\lambda_j\}_{j=1}^{N}]$ (cf. Eq. \eqref{EPBA}), 
with the $\lambda_j$ being rapidities of the ground state itself.
The asymptotic expansions of $K$ and $v_s$ for small and large $\gammap$ are known analytically (e.g. for $\gammap\to\infty$ one has $K=1$ and $v_s=v_F$).
For generic values of $\gammap$, $K$ and $v_s$ must be extracted from the numerical solution of Bethe Ansatz equations \eqref{log-BAE} 
and the evaluation of $v_N$ in Eq.~\eqref{vN}. The result for $K$ is shown in Fig.~\ref{fig:K} and compared with the known asymptotic expansions.

The field theory that underlies the Luttinger liquid Hamiltonian \eqref{luttinger} is the conformal field theory (CFT) of a free massless compact boson in 
two-dimensional euclidean space (see e.g. \cite{DiFrancesco,Gogolin-Tsvelik,m-book}). The action is read off from the Hamiltonian \eqref{luttinger} and it is }
\be\label{boson-cft}
S=\frac{1}{2\pi K}\int \dd\tau \int_0^L \dd x \ \left[ \frac{1}{v_s} (\de_\tau\hat\phi)^2 + v_s (\de_x\hat\phi)^2\right],
\ee
where $\tau$ is the imaginary time. 
\section{One-particle density matrix in a periodic gas}
\label{sec:corr-homo}

The asymptotic behaviour of the one-particle density matrix for the anyonic Lieb-Liniger model \eqref{anyon-Lieb-Liniger} 
can be written in terms of the vertex operators 
\be\label{vertex}
\A_{m,n}(x)= e^{im\hat\phi(x)} e^{in\hat\theta(x)}.
\ee
{thanks to  Eq.~\eqref{bosonization}. Indeed, the correlation}
\be\label{g1}
g_1(x,x^\prime)\equiv\braket{\Psikd(x)\Psik(\xp)},
\ee
admits for $|x|\gg\rho^{-1}$ the low-energy expansion
\be\label{1pdm}
\fl g_1(x,\xp)\simeq \rho \sum_{m=-\infty}^\infty \frac{(b_m^\kappa)^2  \ e^{i(2m+\kappa)\pi \rho (x-\xp)}}{\rho^{\ 2\Delta_{2m+\kappa,1}}}  \braket{\A_{2m+\kappa,-1}(x) \A_{-2m-\kappa,1}(\xp)}_{\mathrm{pbc}}.
\ee
Here we rescaled the amplitudes $B_m^\kappa$ in Eq.~\eqref{bosonization} as
\be\label{b-scaling}
B_m^\kappa=b_m^\kappa(\gammap) \ \rho^{-\Delta_{2m+\kappa,1}},
\ee
where $b^\kappa_m$ are dimensionless coefficients and 
\be\label{scaling}
\Delta_{m,n}=\frac{1}{4}\left(m^2 K + n^2/K \right),
\ee
is the scaling dimension of the vertex operator~\eqref{vertex}. 
The expectation value of the product of vertex operators appearing in Eq.~\eqref{1pdm} is known from CFT for different types of boundary conditions. 
In the case of periodic boundary conditions ({\rm pbc}) one has (see, e.g., Ref.~\cite{Cazalilla2004,m-book})
\be\label{vertex-pbc}
\braket{\A_{m,n}(x) \A_{-m,-n}(\xp)}_{\mathrm{pbc}}=  \frac{e^{i\pi n m \  \sgn(x-\xp)/2}}{\left|\frac{L}{\pi}\sin(\frac{\pi (x-x')}{L})\right|^{2\Delta_{m,n}}}.
\ee
Plugging Eq.~\eqref{vertex-pbc} into Eq.~\eqref{1pdm}, one arrives to \cite{Calabrese2007}
\be\label{1pdm_fin}
g_1(x,0)\simeq \rho \sum_{m=-\infty}^\infty (b_m^\kappa)^2  \ \frac{e^{i(2m+\kappa)\pi \rho x} e^{-i\pi(2 m+\kappa) \sgn(x)/2} }{|N \sin(\pi x/L)/\pi|^{2\Delta_{2m+\kappa,1}}}\,.
\ee
In order to give predictive power to the above sum, we should both identify the leading terms in $m$ and calculate explicitly the non-universal amplitudes 
$b_m^\kappa$. The former issue has been already discussed at length in the literature within the Luttinger liquid approach, see e.g.  \cite{Calabrese2007}. 
In the thermodynamic limit, for large $N$ and at fixed $(x-x')/L$, from Eq. \eqref{1pdm_fin} the leading terms are those with the smallest scaling dimension 
$\Delta_{2m+\kappa,1}$. For bosons, i.e. $\kappa=0$, the leading term is the one with $m=0$ and the first subleading ones are those with $m=\pm1$, that are 
equal because of the symmetry under exchange of $x\leftrightarrow x'$. 
As we move away from $\kappa=0$ (obviously  towards positive $\kappa$), the leading term is always $m=0$, but the harmonic with $m=1$  becomes 
smaller while $m=-1$ increases. At the fermionic point $\kappa=1$, the term $m=-1$ becomes equal and opposite to $m=0$, again by exchange 
statistics. Close to $\kappa=1$, although $m=0$ is the only true  leading term, the one with $m=-1$ is very similar in magnitude and cannot be neglected for distances
large but finite. 

Finally, we need to recall that the harmonic expansion in Eq. \eqref{1pdm_fin} is not an exact expansion. 
Each term in the sum gets anharmonic corrections (descendent fields in CFT) for which each term in the sum is multiplied by a power series in $N^{-1}$.
There are techniques to access these subleading terms but their discussion is much beyond the scope of this paper. 

\begin{figure}[t]
\centering
\includegraphics[width=0.72\textwidth]{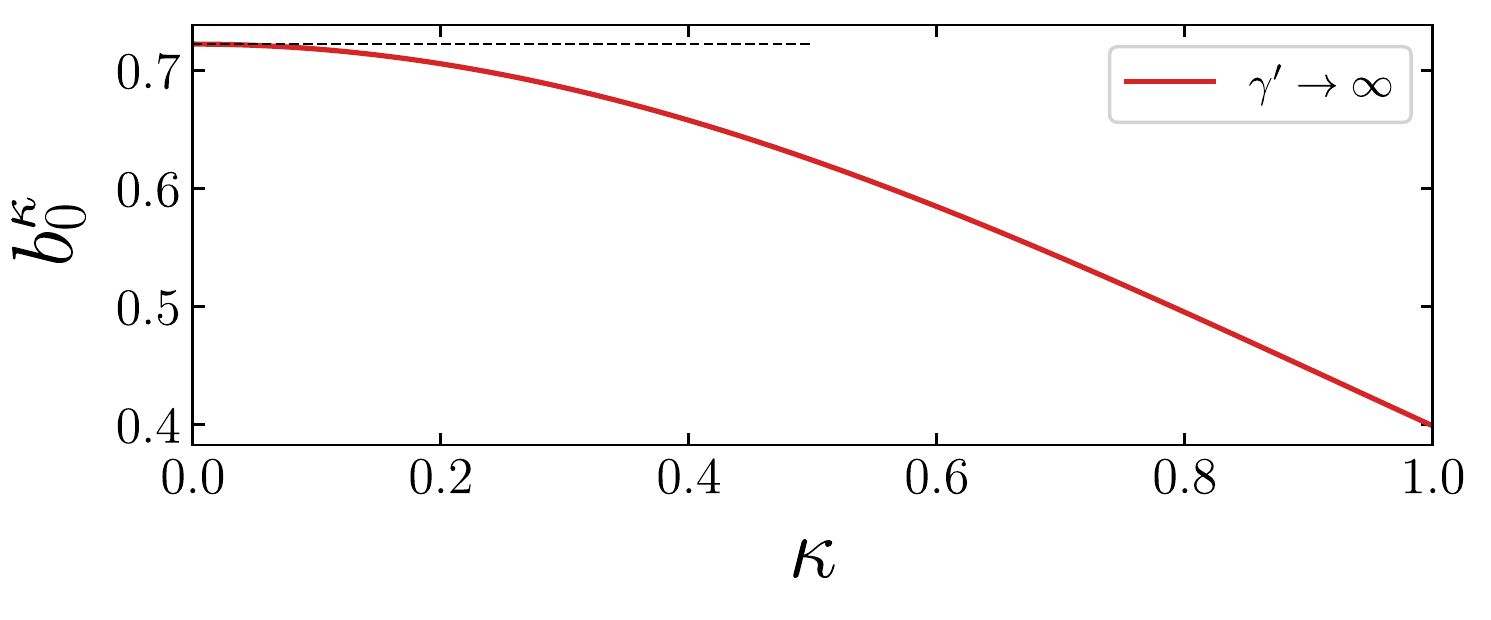}
\includegraphics[width=0.7\textwidth]{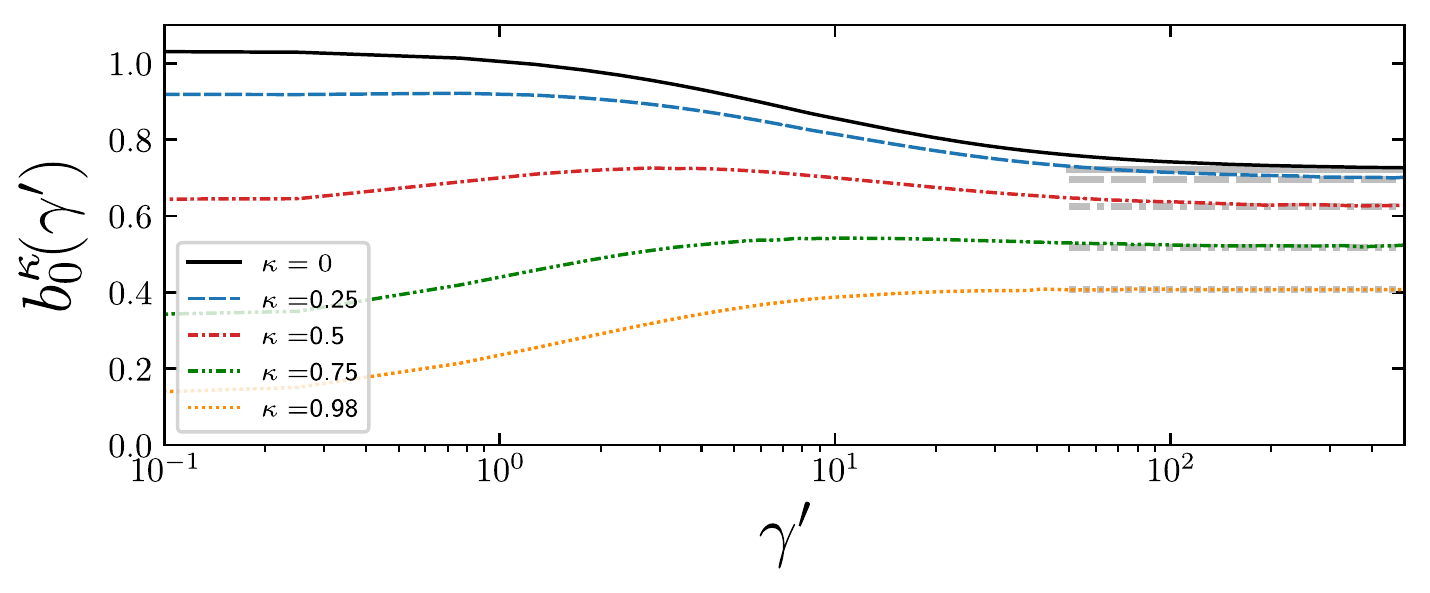}
\caption{The leading amplitude $b_0^\kappa$. 
Top: the limit of strong interactions, $\gammap\to\infty$. Eq.~\eqref{Santachiara} as function of the anyonic parameter $\kappa$ ({\it full line}). 
For $\kappa=0$, one recovers the well-known value ${\mathrm{G}^2(3/2)/(2\pi)^{1/4}}\simeq 0.722$ of impenetrable bosons \cite{Vaida79,Vaida79b} ({\it dashed line}).
Bottom: as a function of $\gamma^\prime$ extracted numerically using \eqref{coef-formula}. 
Different curves show different values of the anyon parameter $\kappa$. 
The ATG results~\eqref{Santachiara} ({\it bold asymptotes}) are recovered for $\gamma^\prime \To\infty$.
}
\label{fig:coef}
\end{figure}

At this point the only missing ingredient is the estimation of the amplitude $b_m^\kappa$. 
In the limit of strong repulsive interactions $\gammap\to\infty$, {aka} anyonic Tonks-Girardeau (ATG) gas \cite{Girardeau06}, 
the leading amplitude $b^\kappa_0$ {is analytically known thanks to a Fisher-Hartwig calculation} \cite{Santachiara2007,Santachiara2008}
\be\label{Santachiara}
\lim_{\gammap\to\infty} b^\kappa_0(\gammap) = {\frac{\mathrm{G}\left(\frac{3+\kappa}{2}\right) \mathrm{G}\left(\frac{3-\kappa}{2}\right)}{(2 \pi)^{\frac{1+\kappa^2}4}}}, 
\ee
with $\mathrm{G}(\cdot)$ the Barnes G-function. The values of $b^\kappa_0(\infty)$ are plotted in Fig.~\ref{fig:coef} (top) as a function of $\kappa$. For $\kappa=0$, Eq.~\eqref{Santachiara} reduces to the well-known result for impenetrable bosons \cite{Vaida79,Vaida79b}. 

For finite values of the interactions, the calculation of $b_m^\kappa$ requires the knowledge of the field form factor, 
which has only been recently obtained  for the anyonic gas \cite{Piroli2020}. We now perform this calculation following the logic of 
Ref.~\cite{Bogoliubov1987} {for the bosonic case} (revisited more recently in Refs.~\cite{SGCI11,SPCI12,Brun2018}).
The first step is the identification, through operator-state correspondence, of the vertex operators $\A_{m,n}$ of the Luttinger liquid with an
excited state $\ket{\{\mu_j\}_{j=1}^{N}}$ of the microscopic model \eqref{anyon-Lieb-Liniger}. 
For sufficiently large system sizes $L\to \infty$, this {identification is unambiguous, as detailed} in the \ref{app:matching}. 
Next, for given $N$ and $L$ such that $N/L=\rho$, we extract the non-universal amplitude $b_m^\kappa$ using the formula (see e.g. \cite{Brun2018}):
\be\label{coef-formula}
b^\kappa_m(\gammap)=\limTh \left[\left(\frac{L}{2\pi\rho}\right)^{\Delta_{2m+\kappa,1}} \ \frac{\braket{\{\lambda_j\}_{j=1}^{N}|\Psikd(0)|\{\mu_j\}_{j=1}^{N-1}}}{\sqrt{\braket{\{\lambda_j\}|\{\lambda_j\}} \ \braket{\{\mu_j\}|\{\mu_j\}}}}\right]\,,
\ee
where $\Delta_{2m+\kappa,1}$ is given in \eqref{scaling}, $\ket{\{\lambda_j\}}$ is the $N$-particle ground state of the anyon Lieb-Liniger model \eqref{anyon-Lieb-Liniger}, and $\ket{\{\mu_j\}}$ is the $(N-1)$-particle excited state of \eqref{anyon-Lieb-Liniger} associated with the vertex operator $\A_{2m+\kappa,-1}$, which is 
explicitly constructed in \ref{app:coefficient}. 
In practice, we determine the value of $b_m^\kappa$ in Eq.~\eqref{coef-formula} for a set of large but finite value of $N$ and $L=N/\rho$ and we extrapolate {to infinite $N$} with a polynomial fit in $1/N$.  Further information and technical details can be found in \ref{app:coefficient}. The result for the leading amplitude $b_0^\kappa$ is shown in Fig.~\ref{fig:coef} (bottom) as a function of $\gamma^\prime$.
%

\begin{figure}[t]
\centering
\includegraphics[width=.9\textwidth]{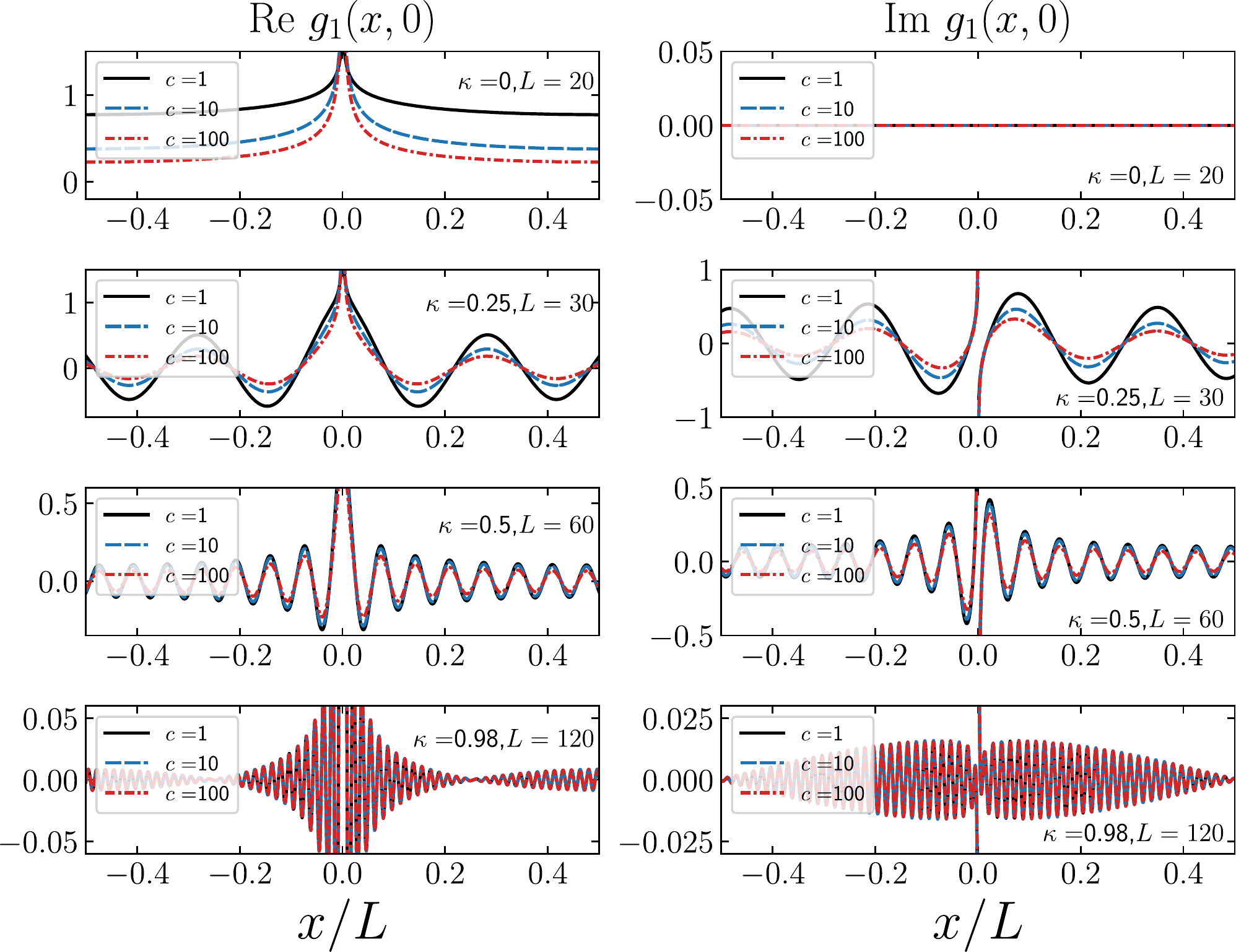}
\caption{The leading order behaviour of the one-particle density matrix \eqref{1pdm_fin} for a system with periodic boundary conditions. 
Different curves show the correlations for $c=1,10,100$ while different panels show the behaviour on varying $\kappa$. 
The data are for $\rho=1$, i.e., $N=L$. The considered value of $L$ is reported in the legend of each plot.} 
\label{fig:corr_pbc}
\end{figure}

Now, we have all ingredients for the explicit evaluation of the correlation function Eq.~\eqref{1pdm_fin}, since we can compute all the amplitudes $b_m^\kappa$.
Here we only show results for the leading term with $m=0$, except close to the fermion point where we also add the harmonic $m=-1$ which has 
a very similar magnitude, as we already stressed.
In Fig.~\ref{fig:corr_pbc} we show the asymptotic results for the real and the imaginary parts of $g_1(x,0)$ for four values of the anyonic parameter and for 
the coupling $c$ going from $1$ to $100$.  
In the strong interaction limit $\gammap\to\infty$, our findings agree with those of Refs.~\cite{Santachiara2007,Calabrese2007,Santachiara2008,Marmorini16}. 
The main qualitative effects of the anyonic statistics are: (i) the presence of oscillations with a frequency that increases with $\kappa$,
(ii) a slow reduction of the peak at $x=x'$ moving from bosons to fermions. 
Both these features are clear from Eq.~\eqref{1pdm_fin} and looking at Fig.~\ref{fig:corr_pbc}. 
For $\kappa\to 1$, we find a beating effect resulting from two oscillations (the harmonics $m=0$ and $m=-1$) with almost the same amplitude that manifests 
as nodes in the {envelopes of the} real and imaginary parts of the one-particle density matrix, in contrast with the monotonic behaviour  observed for smaller $\kappa$. 
However, all these features were already known from the studies in the Tonks-Girardeau regime and remain qualitatively unchanged for finite interaction.
The effect of a finite coupling $c$ is instead investigated here for the first time. 
Fig. \ref{fig:corr_pbc} shows that the correlation is enhanced at large distances with decreasing $c$ for all $\kappa$. 
This effect is more pronounced close to the bosonic point and slowly vanish as we approach fermions.
Indeed, when $\kappa\to 1$, one recovers the case of impenetrable particles independently of the value of $c$ since the effective coupling $c^\prime \to \infty$, cf. Eq. \eqref{cp}.

\begin{figure}[t]
\centering
\includegraphics[width=\textwidth]{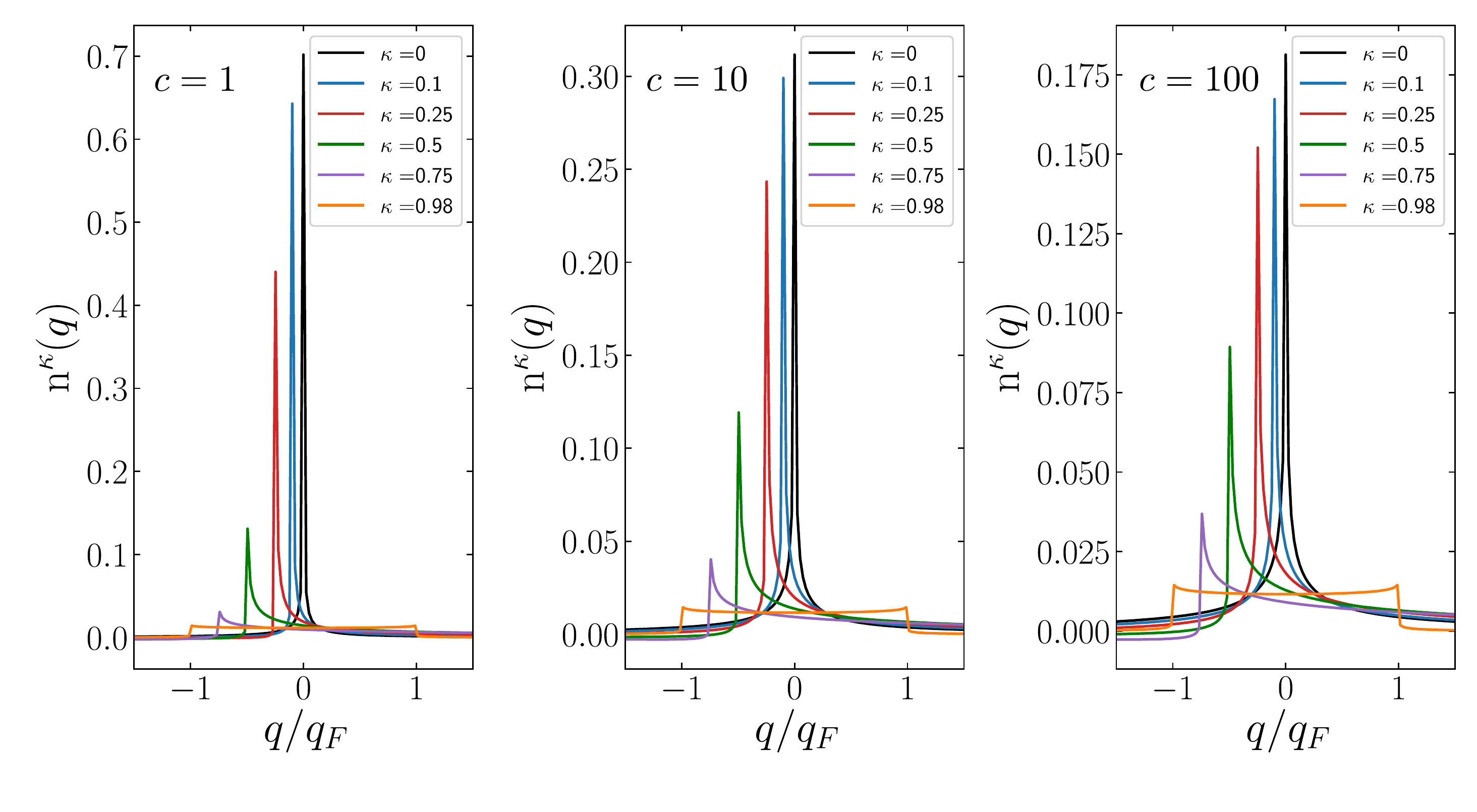}
\caption{Momentum distribution function ${\rm n}^\kappa(q)$ {in Luttinger-liquid approximation} as a function of the rescaled momentum $q/q_F$ {(see the text)}. The different curves show ${\rm n}^\kappa(q)$ for different values of the anyonic parameter $\kappa$ between $0$ and $1$ while the different panels show the behaviour on varying $c=1,10,100$. 
{The data are obtained from the fast Fourier transform of Eq.~\eqref{1pdm_fin}}. 
The plots are made with $N=L=81$. For $c=100$ ({\it right panel}), our findings agree with the ATG results of Ref.~\cite{Santachiara2008}.}\label{fig:mom-distr}
\end{figure}

It is also interesting to investigate the behaviour of the momentum distribution function ${\rm n}^\kappa(q)$ \cite{Santachiara2007,Santachiara2008,Colcelli18b}, defined as the Fourier transform of the one-particle density matrix 
\be
{\rm n}^\kappa(q)=\frac{1}{L}\int_0^L \dd x \ e^{2i\pi q x/L}\ g_1(x,0).
\ee
{Here the {\it integer} $q$ is not the momentum, but the corresponding quantum number. The actual momentum is $k=2\pi q/L$.}
The evaluation of ${\rm n}^\kappa(q)$ can be easily carried out with fast Fourier transform routines applied on data samples of $g_1(x,0)$ with equally spaced points $x$. 
The momentum distribution function obtained through the Luttinger liquid approach is expected to be quantitative accurate only for small values of the momentum
close to $q_F=(N-1)/2$, since this conformal technique is unable to capture the short-distance features of correlation functions. 
The result at leading order is shown in Fig.~\ref{fig:mom-distr} for different values of the interaction strength and on varying $\kappa$ between $0$ and $1$, {plotted as function of $q/q_F$ which is equal to $k/k_F$}. The momentum distribution ${\rm n}^{\kappa=0}(q)$ of bosonic particles exhibits a peak in $q=0$ with height $\sim N^{-1/(2K)}$. {Turning on the anyonic parameter} $\kappa>0$, the peak is dragged backwards to the point $q=-\kappa q_F$, while {the divergence weakens} as the height of the peak changes to $\sim N^{-\alpha(\kappa)}$ with $\alpha(\kappa)=\frac{1}{2}(1/K+ \kappa^2 K)$. This singularity ultimately becomes the discontinuity at $q=-q_F$ when $\kappa\simeq 1$. 
The other discontinuity at $q=q_F$ is instead generated from weaker singularity at $q=(2-\kappa)q_F$  that gets stronger and stronger as $\kappa$ gets close to $1$, up to becoming of leading order when $\kappa=1$ \cite{Santachiara2008}.
We recall that for large $k$, the momentum distribution function for any $\kappa\neq 1$ and arbitrary $c\neq 0$ 
presents a universal tail going like $k^{-4}$ \cite{Santachiara2007}.

\section{Anyonic Lieb-Liniger model in confining potentials}
\label{sec:conf_pot}
Hereafter, we move on to study anyonic gases in inhomogeneous settings. In particular, in this section we adapt to anyons the approach of 
Ref.~\cite{Dubail17} for the characterisation of a Tonks-Girardeau bosonic gas in the presence of arbitrary trapping potentials. 
It relies on the assumption of scales separation \cite{Dubail17,Brun2017,Brun2018,Ruggiero2019,Bastianello2020,j-lec,Y-thesis} and the systematic use of a local 
density approximation (LDA) within fluid cells of mesoscopic length. 

Let us consider an external potential which couples to the density operator of the system. 
The anyonic Lieb-Liniger Hamiltonian~\eqref{anyon-Lieb-Liniger} gets modified as
\bea\label{model}
\hat{H}=\int \dd x \left[ \hat\Psi^\dagger_{\kappa}(x)\left(-\de_x^2-\mu+V(x)\right)\hat\Psi_{\kappa}(x) +c \,\hat\Psi_\kappa^\dagger(x)^2\hat\Psi_\kappa(x)^2\right],
\eea
where we added a chemical potential $\mu$. {Indeed}, from now on, we work at fixed chemical potential $\mu$ {rather than} at fixed particle number $N$. 
However, since our  focus is the ground state of the model \eqref{model}, the two descriptions are equivalent. 
The presence of the trap induces a spatial dependence of thermodynamic quantities and breaks down, in general, the exact solvability of the model discussed in Sec.~\ref{sec:model}.

\begin{figure}[t]
\centering
\includegraphics[width=0.75\textwidth]{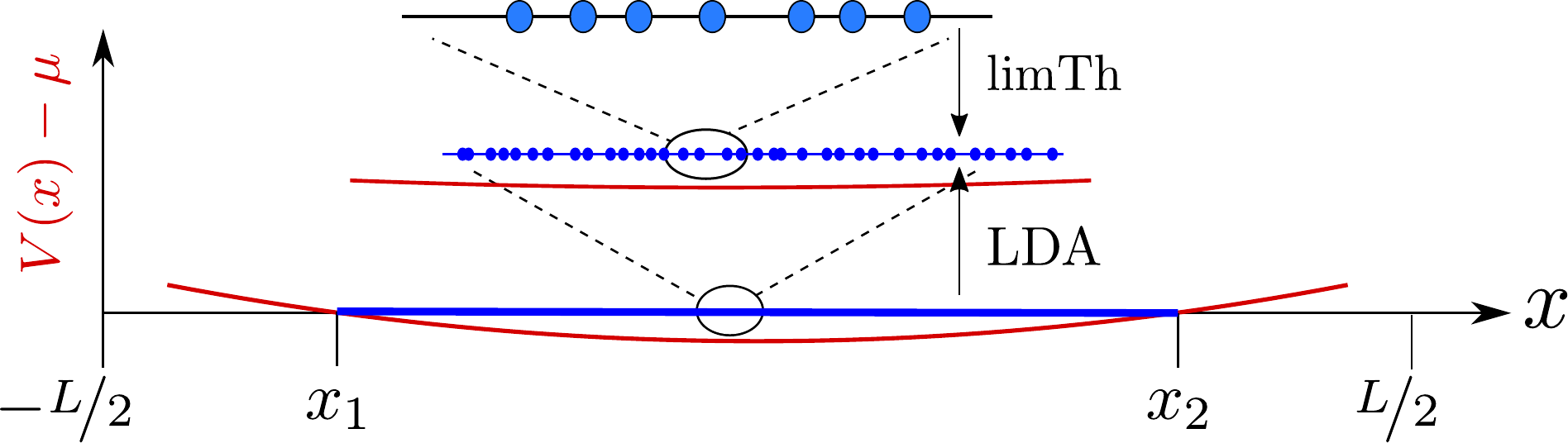}
\caption{Illustration of the separation of scales for a trapped one-dimensional system. We investigate the model \eqref{model} at scales $\ell$ such that $\rho(x)^{-1}\ll \ell \ll \rho(x)|\de_x\rho(x)|^{-1}$, where LDA can be employed self-consistently with TBA.}\label{fig:sep-scales}
\end{figure}

For sufficiently slowly-varying potentials $V(x)$, one can adopt a description of the system over fluid cells of size $\ell$ such that
\be
\rho(x)^{-1} \ll \ell \ll \rho(x) | \partial_x \rho(x)|^{-1},
\label{LDA}
\ee
where $\rho(x)$ is the expectation value of $\hat\rho(x)$ within the fluid cell at position $x$.
On each fluid cell then, the system appears locally homogeneous (because {$\ell$ is smaller than the length over which the density changes $\rho(x) | \partial_x \rho(x)|^{-1}$}) 
but still contains a thermodynamically relevant number of particles (because {$\ell $ is much larger than the mean interparticle distance} $\rho(x)^{-1}$). 
Under these assumptions, the thermodynamic Bethe Ansatz (TBA) results of Sec.~\ref{sec:model} can be applied consistently with LDA, see Fig.~\ref{fig:sep-scales}. 

\begin{figure}[t]
\centering
\includegraphics[width=\textwidth]{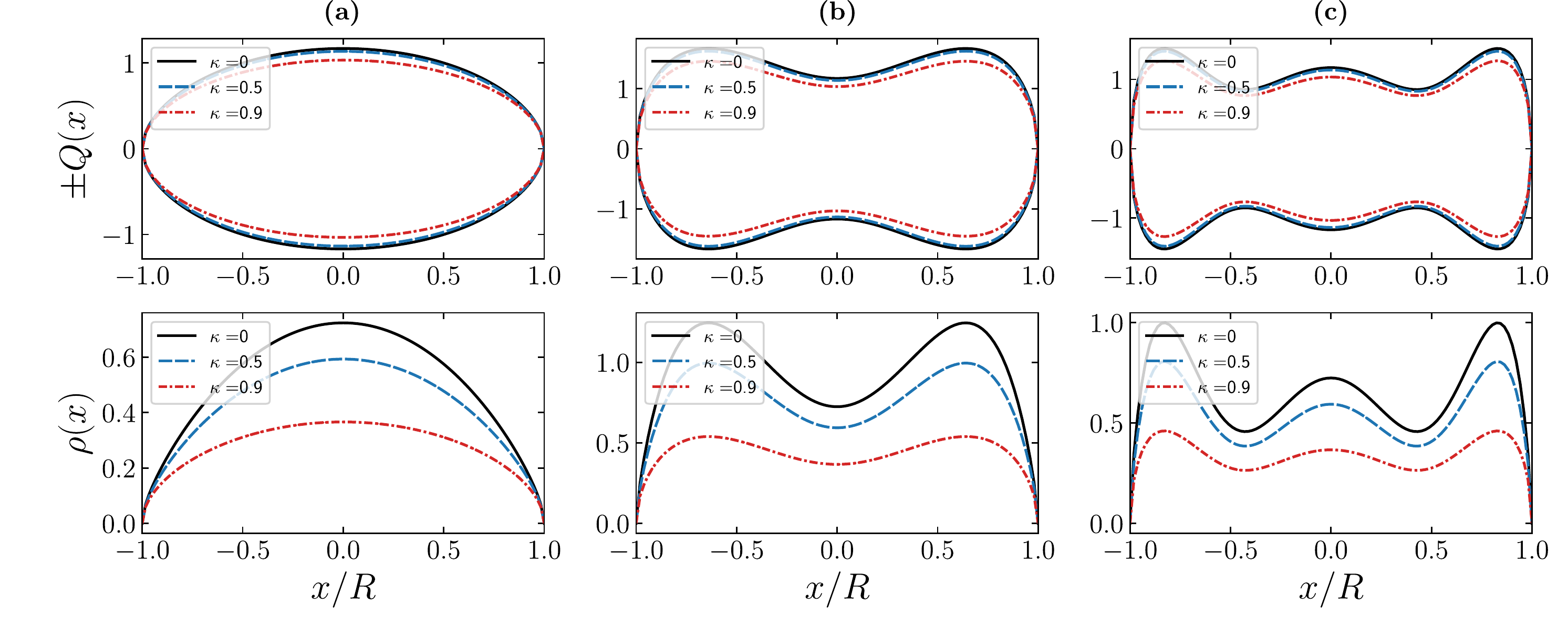}
\caption{({\it Top}) Fermi point spatial distribution $Q_{1,2}(x)$ and ({\it bottom}) the corresponding local particle density $\rho(x)$ for symmetric confining potentials $V(x)=V(-x)$ where $Q_{1,2}=\mp Q$ and $x_{1,2}=\mp R$, as function of the rescaled position $x/R$.
The different curves on each panel show different values of $\kappa$ whereas different $V(x)$ are shown on different columns: {\bf (a)} $V(x)=(2/\LL)^2\,x^2$; {\bf (b)} $V(x)=-(5/\LL)^2\, x^2 +(3.6/\LL)^4\, x^4$; {\bf (c)} $V(x)=(5.3/\LL)^2\, x^2 -(4.8/\LL)^4\, x^4 +(3.6/\LL)^6\, x^6$. Numerical data are obtained setting $c=1$, $\mu=1$ and $\LL=60$. 
}\label{fig:LDA}
\end{figure}

For simplicity, we only consider here confining potentials $V(x)$ and chemical potential $\mu$ such that the effective chemical potential 
$\mu_{\small\mathrm{eff}}(x)= \mu-V(x)$ has exactly two roots, {denoted by} $x_1,x_2$. 
We have $\mu_{\small\mathrm{eff}}(x)>0$ for $x_1< x< x_2$ and negative elsewhere. 
The gas is then confined in the interval $[x_1,x_2]$. Close to $x_1$ and $x_2$ the density vanishes and the separation of scales condition \eqref{LDA}
cannot be satisfied.
The generalisation to multiple roots (with the gas semiclassically confined in disconnected intervals) is straightforward.
The pseudo energy definition~\eqref{pseudo} has to be modified by adding the effective chemical potential.
The local Fermi point $Q_{1,2}(x)$ are obtained within LDA by solving the consistency equation for a given fluid cell at $x\in [x_1,x_2]$
\be\label{LDA-fermi}
\Bigg\{ 
\begin{array}{l}\epsilon(\lambda,x)=\lambda^2 -\mu +V(x)+\frac{1}{2\pi} \int_{{Q_1}(x)}^{{Q_2}(x)} \dd\alpha\,  \frac{2\cp}{c^{\prime\ 2} + (\lambda-\alpha)^2}\, \epsilon(\alpha,x),
 \\[5pt]
 \epsilon({Q_{1,2}}(x),x)=0.
\end{array}
\ee
Similarly, {we introduce the local root density $\rho_p(\lambda,x)$ which satisfies the TBA equation}~\eqref{TBA} for each fluid cell.
Finally the local particle density is 
\be\label{LDA-rho}
\rho(x)=\int_{Q_1(x)}^{Q_2(x)} \dd \lambda \; \rho_p(\lambda,x),
\ee
and the mean number of particles $\overline N= \int \dd x \rho(x)$.
In Fig.~\ref{fig:LDA}, the Fermi point distribution and the corresponding particle density, obtained from the numerical solution of  Eqs.~\eqref{LDA-fermi} 
and \eqref{LDA-rho}, are shown for different confining potentials. 

\subsection{Low-energy description {with an inhomogeneous Luttinger liquid}}
The ideas about separation of scales and LDA machinery can also self-consistently be applied within the Luttinger liquid approach for low energy, 
as e.g. done for the bosonic Lieb-Liniger gas in Refs. \cite{Brun2017,Brun2018}.
The  effective Hamiltonian of the inhomogeneous Luttinger liquid is
\be\label{luttinger2}
\hat{H}_{LL}=\frac{1}{2\pi} \int \dd x\,  v_s(x) \, \left(K(x)\,(\de_x\hat\theta)^2 +\frac{1}{K(x)}(\de_x\hat\phi)^2\right),
\ee
which generalises the Luttinger liquid Hamiltonian \eqref{luttinger} to inhomogeneous setups, {allowing for a spatial dependence of both the Luttinger parameter $K$
and the sound velocity $v_s$} \cite{Maslov95,Gangart04,Dubail17,Eisler17,j-lec,Y-thesis,Brun2017,Brun2018,Bastianello2020}. 
{These Hamiltonian parameters depend on $x$ through the spatially varying effective coupling $\gammap(x)\equiv c'/\rho(x)$ (we stress
that the prime is not a derivative here).
Hence, we have $v_s(x)=v_s(\gammap(x))$ and $K(x)=K(\gammap(x))$.}
For a given fluid cell at position $x$ (i.e., for a given value of $\gammap(x)$), the 
local {values of $v_s$ and $K$  are obtained with the techniques of Sec.~\ref{sec:luttinger-liquid}}. 

\begin{figure}[t]
\centering
\includegraphics[width=0.8\textwidth]{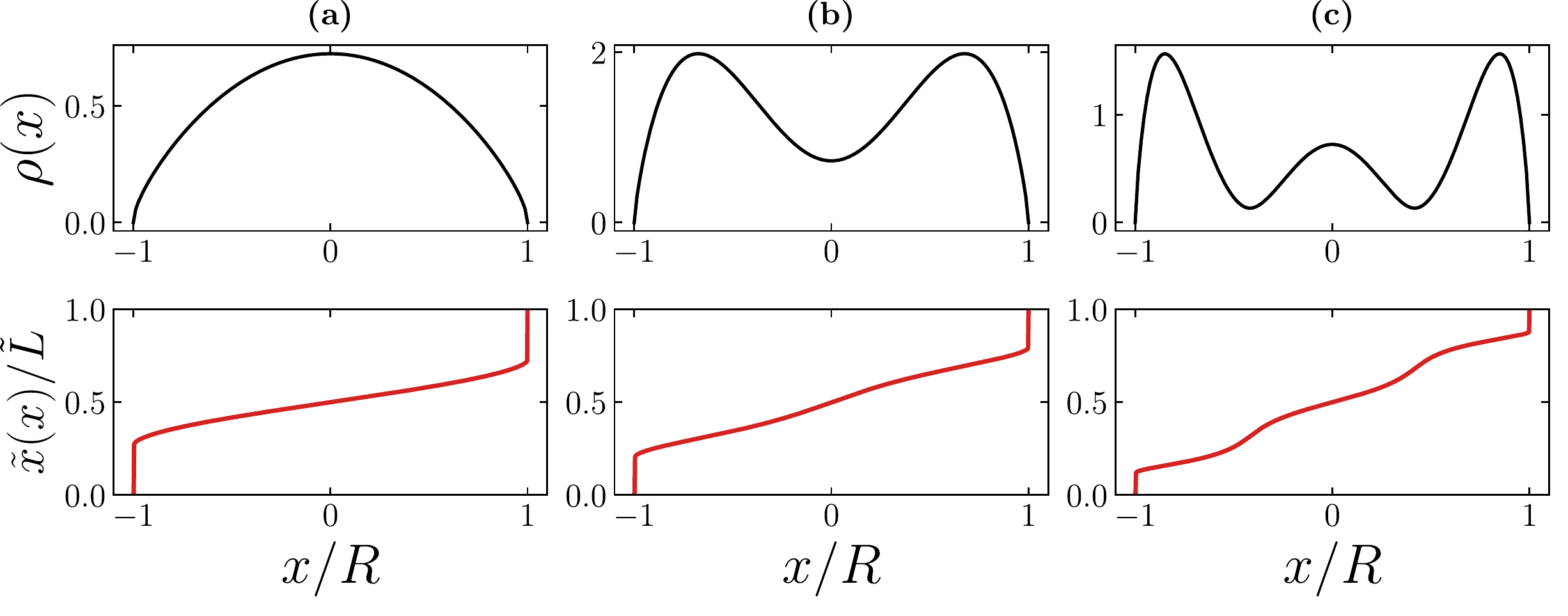}
\caption{({\it Bottom}) The isothermal coordinate $\tilde{x}(x)$ in \eqref{isothermal} and ({\it top}) the corresponding particle density for different confining potentials. In particular, we have set $c=1$, $\mu=1$, $\kappa=0$, $\LL=80$ and, on different columns: {\bf(a)} $V(x)=(2/\LL)^2 x^2$; {\bf(b)} $V(x)=-(7/\LL)^2 x^2 +(4/\LL)^4 x^4$; {\bf(c)} $V(x)=(8/\LL)^2x^2-(6/\LL)^4 x^4 +(4.2/\LL)^6 x^6$. }\label{fig:iso-coord}
\end{figure}

The field theory that describes the properties of the effective Hamiltonian \eqref{luttinger2} is a free massless compact boson with a space-dependent coupling and equipped with a non-flat metric tensor $g_{ab}$ \cite{Y-thesis,j-lec,Dubail17,Eisler17,Brun2017,Brun2018,Bastianello2020}, {with Euclidean action}
\be\label{action}
S=\frac{1}{2\pi} \int \frac{\sqrt{g}\, \dd^2\x}{K(\x)} \, g^{ab}\, (\de_a\hat\phi)(\de_b\hat\phi),
\ee
where $\x=(x,\tau)$ and $g_{ab}$ is the two-dimensional euclidean metric with line element
\be\label{metric-g}
\dd s^2 = g_{ab}\, \dd\x^a\dd\x^b = \dd x^2 + v_s(x)^2 \ \dd\tau^2.
\ee
The non-flat metric $g_{ab}$ can be {eliminated with} the following change of coordinates
\be\label{isothermal}
\tilde{x}(x)\equiv \int_{x_1}^x \frac{\dd y}{ v_s(y)},
\ee
where the new variable $\tilde{x}(x)$ lives in the interval $\tilde{x}(x)\in[0,\tilde{L}]$ with
\be
\tilde{L}=\tilde x(x_2)=\int_{x_1}^{x_2}  \frac{\dd y}{ v_s(y)}.
\ee
The coordinate $\tilde{x}$ physically represents the time needed by a signal emitted from the left boundary $x_1$ to reach the position $x$  
{traveling with velocity $v_s(x)$}. 
It is then easy to see that the change of coordinates $x\to \tilde{x}(x)$ is {\it isothermal}, i.e., it sets the metric \eqref{metric-g} in the diagonal form $\dd s^2= v_s(x)^2\ (\dd\tilde{x}^2 +\dd\tau^2)$. Examples of isothermal coordinates $\tilde{x}$ for different confining potentials are shown in Fig.~\ref{fig:iso-coord}.

{The isothermal coordinate \eqref{isothermal} cancels the local sound velocity in the action~\eqref{action}, leading to}
\be\label{action-isothermal}
S=\frac{1}{2\pi} \int \frac{\dd^2\x}{K(\x)}\,  (\nabla_\x \hat\phi)^2,
\ee
where we have conveniently re-defined $\x=(\tilde{x}(x),\tau)$. 
No change of coordinates {can} instead remove the dependence on $K(\x)$ in Eq.~\eqref{action-isothermal}.
{We stress that the presence of $K(\x)$ also breaks conformal invariance}. 

Note that, in the strong interaction regime $\gammap\to\infty$, the {the Luttinger parameter does not depend on $x$ anymore} and the action \eqref{action-isothermal} 
reduces to the CFT one in Eq.~\eqref{boson-cft} with $K=1$, whose correlation functions are known and can be readily used for our purposes, see Sec.~\ref{sec:corrATG}. 
However, for the generic case with $\gamma^\prime(x)$ finite, standard CFT results are no longer {useful} and the computation of the correlation functions relies instead on the numerical evaluation of the Green's functions of a generalised Laplace operator $\nabla_\x\, \frac{1}{K(\x)}\, \nabla_\x$, see Sec.~\ref{sec:corr-finite}.

\section{One-particle density matrix in {a trapping potential}}
\label{sec:opf_in_setup}

In this section, we build upon the framework introduced in the previous section and finally present our results for the anyonic correlation functions in a trapping potential.
The harmonic expansion of a uniform anyonic field in Eq.~\eqref{bosonization}, readily generalises  to the inhomogeneous setting as 
\be\label{bosonization2}
\Psikd(x) = \sqrt{\rho(x)}\sum_{m=-\infty}^{\infty} B^\kappa_m(x) \, e^{i\pi (2m+\kappa) \int_{x_1}^x \dd y \,\rho(y)} \, e^{-i\hat\theta(x)}\, e^{i(2m+\kappa)\hat\phi(x)},
\ee
where the local non-universal amplitude $B^\kappa_m(x)$ depends on the spatial positions only  through the ratio $\gammap(x)=\cp/\rho(x)$ as 
$B^\kappa_m(x)=B_m^\kappa(\gammap(x))$. In Sec. ~\ref{sec:corr-homo} we detailed how to derive $B_m^\kappa(\gammap)$ for fixed $\gammap$.

Within the separation of scale assumption in Eq. \eqref{LDA}, the asymptotic behaviour of the one-particle density matrix \eqref{g1} is
\bea\label{1pdm-inh}
g_1(x,x^\prime)=&\sqrt{\rho(x) \rho(\xp)} \sum_{m,m^\prime=-\infty}^\infty  e^{i\pi\left[(2m+\kappa)\int_{x_1}^x \dd y \, \rho(y) - (2m^\prime+\kappa)\int_{x_1}^{\xp} \dd y \rho(y)\right]}
\nonumber \\
&\times B^\kappa_m(x)\ B^{\kappa}_{m^\prime}(\xp) \, \braket{\A_{2m+\kappa,-1}(x) \A_{-2m^\prime-\kappa,1}(\xp)}^{\mathrm{curv}}_{\mathrm{obc}},
\eea
which directly follows from Eq.~\eqref{bosonization2}. In this expression, the correlation of vertex operators is computed on a curved space with open boundary conditions (obc). 
{Notice, as a first important difference with the uniform case, that we cannot use translational invariance to remove one of the two sums in  Eq.~\eqref{1pdm-inh}.
The first step to evaluate the correlation \eqref{1pdm-inh} is to perform a} Weyl transformation $g\to e^{2\sigma(x)} g$ with line element 
\be
\dd s^2= \dd x^2 + v_s(x)^2 \ \dd \tau^2=e^{2\sigma(x)}(\dd\tilde{x}^2 +\dd\tau^2),
\ee
where $e^{\sigma(x)}\equiv v_s(x)$ and $\tilde{x}$ given in Eq.~\eqref{isothermal}. 
Vertex operators behave as primary fields
\be 
\A_{m,n}(x)=(v_s(x))^{-\Delta_{m,n}(x)}\, \A_{m,n}(\tilde{x}(x)),
\ee
with spatially varying scaling dimension
\be
\Delta_{m,n}(x)=\frac{1}{4}\left(m^2 K(x) +\frac{n^2}{K(x)}\right).
\ee
{The correlation of vertex operators in Eq.~\eqref{1pdm-inh} can be then rewritten in terms of the one} on a flat geometry with isothermal coordinates as
\bea\label{weyl}
\braket{\A_{m,n}(x)\A_{-m^\prime,-n}(x^\prime)}_{\mathrm{obc}}^{\mathrm{curv}}=&\left(v_s(x)\right)^{-\Delta_{m,n}(x)} \ \left(v_s(\xp)\right)^{-\Delta_{-m^\prime,-n}(\xp)}
\nonumber\\&\times 
\braket{\A_{m,n}(\tilde{x}(x)) \A_{-m^\prime,-n}(\tilde{x}(x^\prime))}_{\mathrm{obc}}^{[K]}.
\eea
The notation $\braket{\cdot}^{[K]}_{\mathrm{obc}}$ remarks that the expectation value on the r.h.s. of Eq.~\eqref{weyl}  is taken on a flat geometry but in 
{an inhomogeneous medium with spatially varying Luttinger parameter $K(\tilde{x})$}.

\begin{figure}[t]
\centering
\includegraphics[width=.9\textwidth]{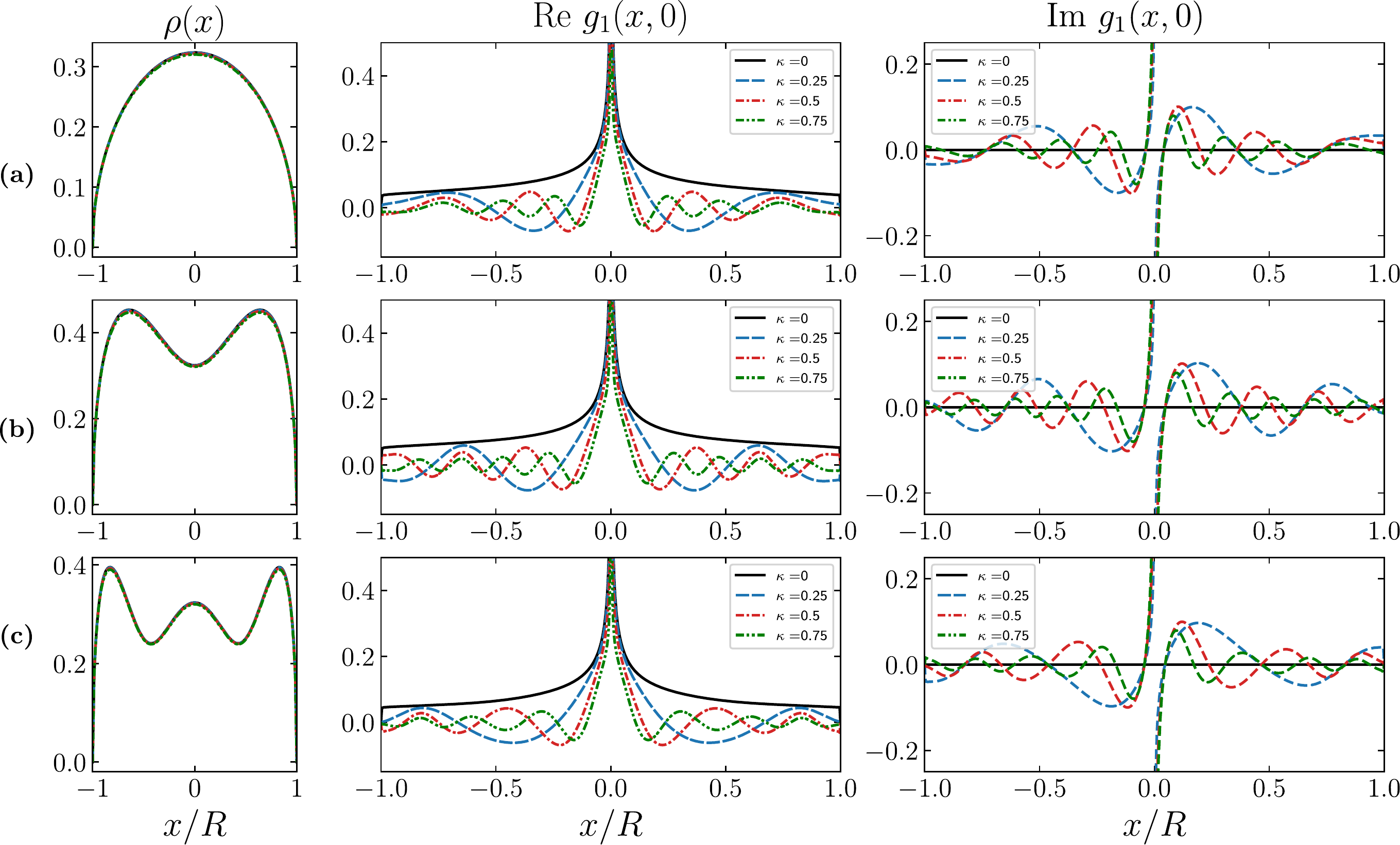}
\caption{({\it Right and middle}) The asymptotic behaviour of the one-particle density matrix \eqref{1pdm_general} for the ATG gas and ({\it left}) the corresponding particle density as functions of the rescaled position $x/R$, with $x_{1,2}=\mp R$. The different curves on each panel show different values of $\kappa$ whereas different confining potential $V(x)$ are shown on different rows: (a) $V(x)=(2/\LL)^2\,x^2$; (b) $V(x)=-(5/\LL)^2\, x^2 +(3.6/\LL)^4\, x^4$; (c) $V(x)=(5.3/\LL)^2\, x^2 -(4.8/\LL)^4\, x^4 +(3.6/\LL)^6\, x^6$.  Data are obtained setting  $c=50$, $\mu=1$, $\LL=80$
}
\label{fig:TG_corr}
\end{figure}

Plugging Eq.~\eqref{weyl} into Eq.~\eqref{1pdm-inh}, we arrive at our main result for the one-particle density matrix of the trapped anyonic Lieb-Liniger gas \eqref{model}
\bea
\label{1pdm_general}\fl
g_1(x,x^\prime)=&\sum_{m,m^\prime=-\infty}^\infty  e^{i\pi\left[(2m+\kappa)\int_{x_1}^x \dd y \, \rho(y) - (2m^\prime+\kappa)\int_{x_1}^{\xp} \dd y \rho(y)\right]} \ \frac{b^\kappa_m(x)\ \sqrt{\rho(x)}}{\left[\rho(x)\, v_s(x)\right]^{\Delta_{2m+\kappa,1}(x)}}
\nonumber \\
\fl & \times \frac{b^\kappa_{m^\prime}(\xp)\ \sqrt{\rho(\xp)}}{\left[\rho(\xp)\, v_s(\xp)\right]^{\Delta_{2m^\prime+\kappa,1}(\xp)}} \ \braket{\A_{2m+\kappa,-1}(\tilde{x}(x)) \A_{-2m^\prime-\kappa,1}(\tilde{x}(x^\prime))}_{\mathrm{obc}}^{[K]},
\eea
where, as in the homogeneous case, we have set $b_m^\kappa(x)=B_m^\kappa(x)\ \rho(x)^{\Delta_{2m+\kappa,1}(x)}$.
Each term in the sum can be exactly calculated (numerically) with known techniques, as we will explain and do in the following. 
Even here we completely ignore the contribution of anharmonic terms  (descendent fields).

\subsection{Anyonic Tonks-Girardeau trapped gas}\label{sec:corrATG}
In the Tonks-Girardeau regime $\gamma^\prime(x)\To\infty$, we have a uniform Luttinger parameter $K(x)\equiv 1$ and 
we can employ standard boundary CFT techniques \cite{DiFrancesco} for the correlation functions of vertex operators with obc. 
The final result for the vertex correlation function is (see, e.g., \cite{DiFrancesco,Cazalilla2004})
\bea\label{cazalilla}\fl
\braket{\A_{m,n}(\tilde{x}) \A_{-m^\prime,-n}(\tilde{x}^\prime)}_{\mathrm{obc}}^{[1]}&=\left[\mathrm{d}(2\tilde{x},2\tilde{L})\right]^{-\frac{1}{4}(m^2-n^2)} \ \left[\mathrm{d}(2\tilde{x}^\prime,2\tilde{L})\right]^{-\frac{1}{4}(m^{\prime \ 2}-n^2)} \nonumber\\ \fl
&\qquad\times \frac{ \left[\mathrm{d}(\tilde{x}+\tilde{x}^\prime,2\tilde{L})\right]^{\frac{1}{2}(m m^\prime-n^2)}}{\left[\mathrm{d}(\tilde{x}-\tilde{x}^\prime,2\tilde{L})\right]^{\frac{1}{2}(m m^\prime+n^2)} } \ e^{i\pi n(m+m^\prime) \sgn(\tilde{x}-\tilde{x}^\prime)/4},
\eea
with shorthands $\tilde{x}=\tilde{x}(x)$, $\tilde{x}^\prime=\tilde{x}(x^\prime)$  and $\mathrm{d}(\tilde{x},\tilde{L})\equiv \tilde{L} |\sin(\pi \tilde{x}/\tilde{L})|/\pi$. 

\begin{figure}[t]
\centering
\includegraphics[width=\textwidth]{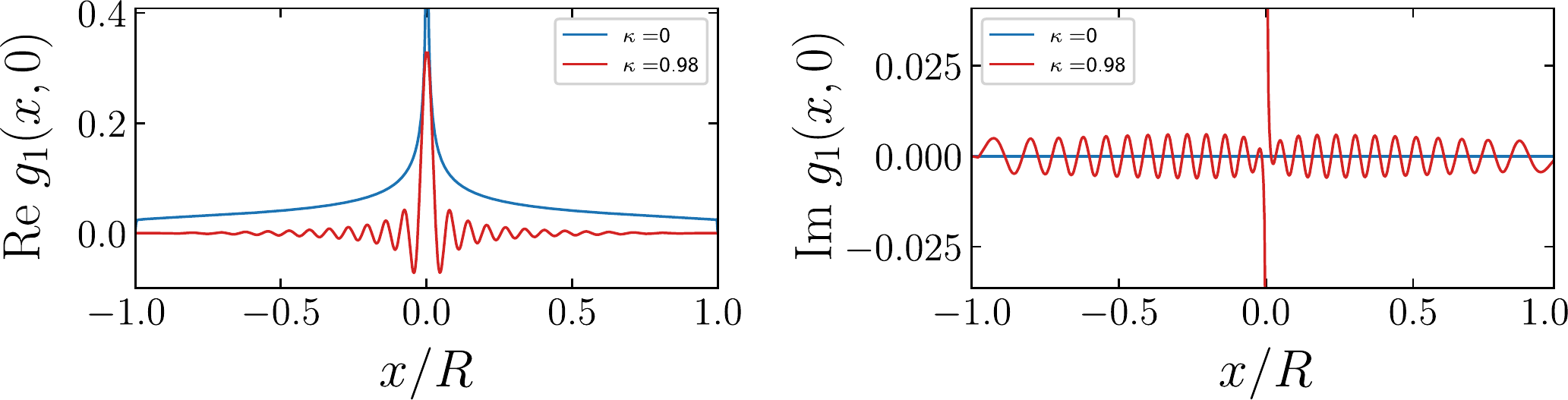}
\caption{The leading term of the one-particle density matrix \eqref{1pdm_general} in the ATG regime for a harmonic trap potential as functions of the rescaled position $x/R$, with $x_{1,2}=\mp R$. In this figure, we plot the bosonic case (corresponding to $\kappa=0$) and  the fermionic one (corresponding to $\kappa=0.98\simeq 1$). Numerical data are obtained setting   $c=50$, $\mu=1$  and $V(x)=(2/\LL)^2 x^2$, $\LL=200$.
}\label{fig:TG-fermi}
\end{figure}

Plugging Eq. \eqref{cazalilla} into Eq.~\eqref{1pdm_general} and exploiting the knowledge of $b_m^\kappa$ (cf. Eq.~\eqref{Santachiara} for $b_0^\kappa$), 
we  get the {harmonic expansion} of the one-particle density matrix in the Tonks-Girardeau regime for arbitrary trap potentials. 
In Fig.~\ref{fig:TG_corr} we show the leading behaviour for different confining potential upon varying $\kappa$ (with only $m=m'=0$ at small $\kappa$ 
and adding also $m,m'=-1$ close to the fermion point). 
As one can see in Fig.~\ref{fig:TG_corr}, for $\kappa=0$ there is the typical peaked function of bosons \cite{Brun2017} while, 
increasing $\kappa$, the one-particle density matrix develops oscillations with increasing frequency. 
This tendency ultimately leads for $\kappa\to 1$ to a {\it sinc-like} function arising from the superposition of the two harmonics $m,m'=0,-1$, see Fig.~\ref{fig:TG-fermi}.
Importantly, for a harmonic trapping potential, Eq. \eqref{cazalilla} reproduces the known exact solution of the ATG model \cite{Marmorini16}.
The results for arbitrary trapping potential instead appear here for the first time.

%

\subsection{Finite interaction case}\label{sec:corr-finite}
In this subsection it convenient to work with the normalised isothermal coordinates $r\equiv \tilde{x}/\tilde{L}\in [0,1]$ which induce the rescaling of the vertex operators
\be
\A_{m,n}(\tilde{x}) = (\tilde{L}/\pi)^{-\Delta_{m,n}(r)} \ \A_{m,n}(r).
\ee
For finite interactions $\gamma^\prime(x)$, {the evaluation of the correlation function of vertex operators appearing in the} one-particle density matrix \eqref{1pdm_general} 
is based on the numerical computation of {the three independent} Green's functions of the inhomogeneous Luttinger liquid Hamiltonian \eqref{luttinger}
\bea\label{Gpp}
G_{\phi\phi}(r,r^\prime)\equiv \braket{\hat\phi(r)\ \hat\phi(r^\prime)}_{\mathrm{obc}}^{[K]},\\
\label{Gtt}
G_{\theta\theta}(r,r^\prime)\equiv \braket{\hat\theta(r)\ \hat\theta(r^\prime)}_{\mathrm{obc}}^{[K]},\\
\label{Gpt}
G_{\phi\theta}(r,r^\prime)\equiv \braket{\hat\phi(r)\ \hat\theta(r^\prime)}_{\mathrm{obc}}^{[K]}.
\eea
These Green functions can be calculated in several manners. Here we follow the algorithm developed in the recent paper \cite{Bastianello2020}, see \ref{app:green} for a detailed explanation.
The two-point correlators in Eqs.~\eqref{Gpp} and \eqref{Gtt} are singular for $r\To r^\prime$. 
It is therefore necessary to introduce the regularised Green's functions \cite{Brun2018}
\bea\label{Gpp-reg}
G^\reg_{\phi\phi}(r)=\lim_{r\To r^\prime}\left[G_{\phi\phi}(r,r^\prime)+\frac{1}{4}K(r)\log|r-r^\prime|^2\right]  ,\\
\label{Gtt-reg}
G^\reg_{\theta\theta}(r)=\lim_{r\To r^\prime}\left[G_{\theta\theta}(r,r^\prime)+\frac{1}{4{K(r)}}\log|r-r^\prime|^2\right]\ ,
\eea
so that the second term on the r.h.s. of Eqs.~\eqref{Gpp-reg} and \eqref{Gtt-reg} cancels the divergence of the Green's functions when $r\to r^\prime$, 
see \ref{app:exact} for details.

\begin{figure}
	\centering
	\includegraphics[width=\textwidth]{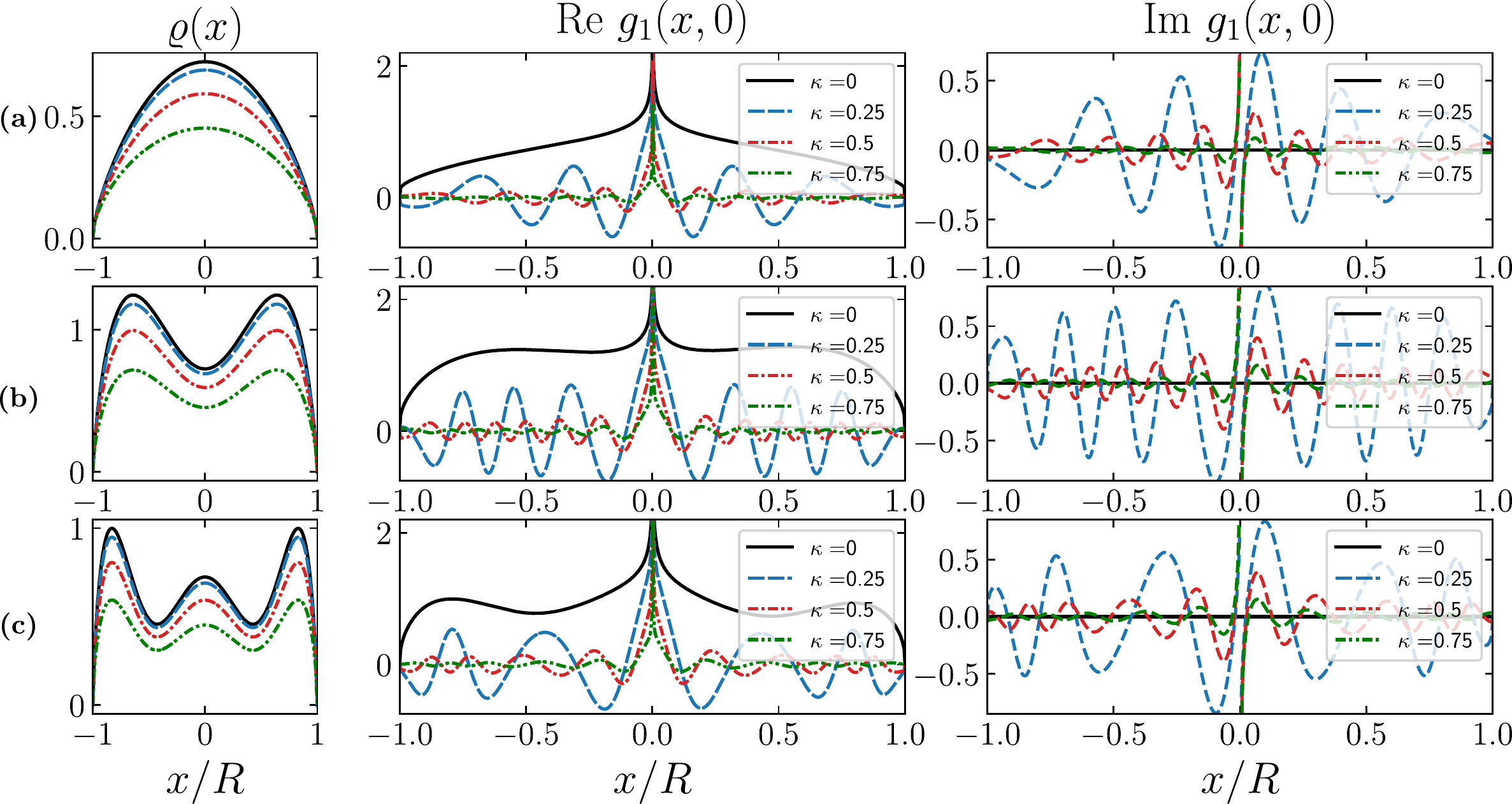}
	\caption{({\it Right and middle}) The leading order behaviour of the one-particle density matrix $g_1(x,0)$ \eqref{1pdm_general} at finite interactions $\gamma^\prime(x)$ and ({\it left}) the corresponding particle density as function of the rescaled position $x/R$, with $x_{1,2}=\mp R$. Figures are obtained with the numerical evaluation of the Green's function \eqref{Gpp}-\eqref{Gpt} (see Eq.~\eqref{inh_vertex} and \ref{app:green}) and of $b^\kappa_0$ (see Fig.~\ref{fig:coef} and \ref{app:coefficient}). 
The different curves on each panel show different values of $\kappa$ whereas different confining potential $V(x)$ are shown on different rows.
{\bf (a)} $V(x)=(2/\LL)^2\,x^2$; {\bf (b)} $V(x)=-(5/\LL)^2\, x^2 +(3.6/\LL)^4\, x^4$; {\bf (c)} $V(x)=(5.3/\LL)^2\, x^2 -(4.8/\LL)^4\, x^4 +(3.6/\LL)^6\, x^6$. Numerical data are obtained setting $c=1$, $\mu=1$ and $\LL=80$.}
	\label{fig:corr-final}
\end{figure}

\begin{figure}
	\centering
	\includegraphics[width=0.85\textwidth]{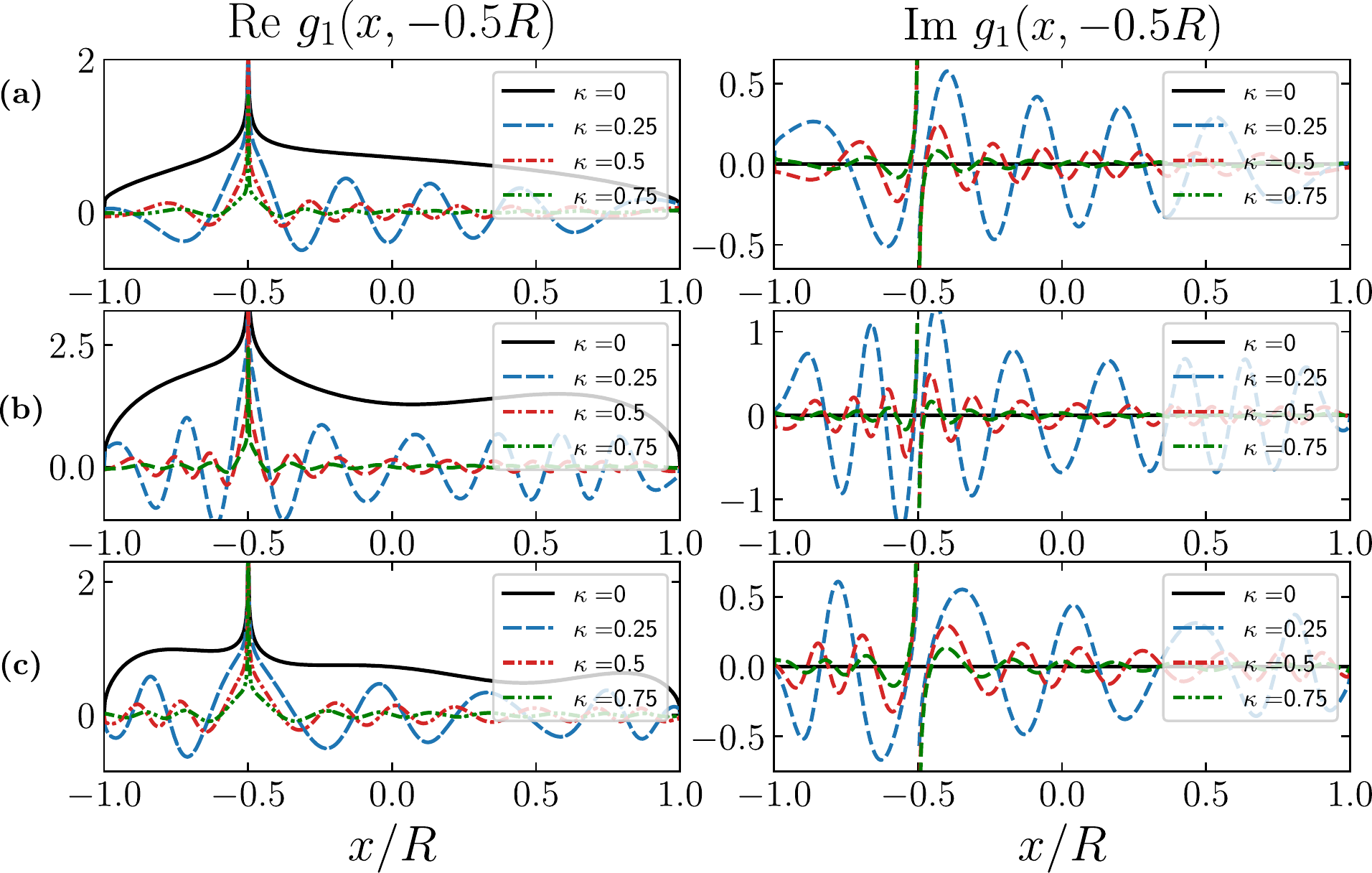}
	\caption{The same as in Fig.~\ref{fig:corr-final} with $x^\prime=-0.5 R$ rather than $0$. 
	}\label{fig:corr-final2}
\end{figure}

Finally Wick theorem can be employed to write the expectation value of vertex operators in Eq.~\eqref{1pdm_general} in terms of the Green functions as \cite{Brun2018}
\be\label{inh_vertex}
\braket{\A_{m,n}(r)\ \A_{-m^\prime,-n}(r^\prime)}_{\mathrm{obc}}^{[K]}={\cal V}^{\phi\phi}_{m,m^\prime}(r,r^\prime) \ {\cal V}^{\theta\theta}_{n}(r,r^\prime) \ {\cal V}^{\phi\theta}_{n,m \atop m^\prime}(r,r^\prime),
\ee
where
\be
{\cal V}^{\phi\phi}_{m,m^\prime}(r,r^\prime) =\exp\left[-\frac{m^2}{2}\ G^\reg_{\phi\phi}(r)-\frac{m^{\prime  2}}{2} \  G^\reg_{\phi\phi}(r^\prime) +m m^\prime \ G_{\theta\theta}(r,r^\prime)\right],
\ee
\be\label{Vtt}
{\cal V}^{\theta\theta}_{n}(r,r^\prime)=\exp\left[-\frac{n^2}{2}\left(G^\reg_{\theta\theta}(r)+G^\reg_{\theta\theta}(r^\prime)\right) + n^2 \  G_{\theta\theta}(r,r^\prime)\right],
\ee
and
\be\fl
{\cal V}^{\phi\theta}_{n,m \atop m^\prime}(r,r^\prime)=
\exp \left(n\left[(m+m^\prime) G_{\phi\theta}(r,r^\prime)  -m G_{\phi\theta}(r,r) - m^\prime G_{\phi\theta}(r^\prime,r^\prime)\right]\right).
\ee
Combining Eqs.~\eqref{inh_vertex} and~\eqref{1pdm_general}, we {get our final} expression for the one-particle density matrix of the trapped anyonic Lieb-Liniger 
model \eqref{model} with finite interaction strength
\bea\label{1pdm_final}\fl
g_1(x,x^\prime)=&\sum_{m,m^\prime=-\infty}^\infty  e^{i\pi\left[(2m+\kappa)\int_{x_1}^x \dd y \, \rho(y) - (2m^\prime+\kappa)\int_{x_1}^{\xp} \dd y \rho(y)\right]} \ \frac{b^\kappa_m(x)\ \sqrt{\rho(x)}}{[\tilde{L}\rho(x) v_s(x)/\pi]^{\Delta_{2m+\kappa,1}(x)}}
\nonumber \\ \fl
 & \times \frac{b^\kappa_{m^\prime}(\xp) \sqrt{\rho(\xp)}}{[\tilde{L}\rho(\xp) v_s(\xp)/\pi]^{\Delta_{2m^\prime+\kappa,1}(\xp)}} 
 {\cal V}^{\phi\phi}_{2m+\kappa,2m^\prime+\kappa}(r,r^\prime)   {\cal V}^{\theta\theta}_{-1}(r,r^\prime)  {\cal V}^{\phi\theta}_{-1,2m+\kappa \atop 2m^\prime+\kappa}(r,r^\prime).
\eea
Each term in the above double sum can be readily worked out numerically putting together all techniques we outlined throughout this paper. 
As topical examples in Figs.~\ref{fig:corr-final} and ~\ref{fig:corr-final2} we report only the leading term in the sum (with $m=m^\prime=0$) 
for three different trapping potentials and for four values of the statistical parameter $\kappa$ (sufficiently far from the fermionic point where 
the mode with $m,m^\prime=-1$ cannot be neglected). 
We report our results for $g_1(x,x^\prime)$ as a function of $x$ at fixed $x^\prime$, with $x^\prime=0$ in Fig.~\ref{fig:corr-final} and $x^\prime=-0.5 R$ in Fig.~\ref{fig:corr-final2}.
Among the various $\kappa$, we also reported the bosonic case with results that perfectly match the ones in Refs. \cite{Brun2017,Brun2018} 
(these have been also tested against accurate density matrix renormalisation group (DMRG) simulations, showing an excellent agreement).
In the figures we can observe that a non-zero anyonic parameter causes strong oscillations of $g_1(x,x')$ as a function of both space variables, that are 
not present for bosons. As a very important difference with the homogeneous case (cf. Fig. \ref{fig:corr_pbc}), these oscillations are not uniform 
and get modulated with the positions.  Our approach is able to capture the fine details of this modulation.
It would be extremely interesting to test quantitatively some of our predictions in numerical simulations, e.g. against DMRG calculations for a 
dilute anyon Hubbard model already performed in the uniform case in Ref. \cite{col}. 

\section{Conclusions}
\label{sec:conclusions}

In this work we combined Luttinger liquid techniques and the exact knowledge of the non-universal amplitudes from Bethe Ansatz to access
the zero-temperature one-particle density matrix of the anyonic Lieb-Liniger model, for finite values of the interactions and different trapping potentials. 
In the homogeneous case, we exploited the Bethe Ansatz formula for the field form factors {recently} derived in Ref.~\cite{Piroli2020} 
to extract numerically non-universal coefficients appearing in the formal expansions for the correlation functions. 
We explicitly computed the one-particle density matrix for finite values of the interactions, thus extending standard results obtained in the Tonks-Girardeau limit. 
In the case of inhomogeneous density profiles, we {generalised to anyonic statistics the techniques of non-uniform Luttinger liquid developed in 
Refs.~\cite{Brun2017,Brun2018,Y-thesis,j-lec,Bastianello2020} for bosons. 
This generalization allowed us to obtain an exact asymptotic expansion for the one-particle density matrix in an arbitrary trapping potential that we 
explicitly worked out for some specific cases, highlighting the main differences arising with respect to the bosonic statistics.}\\

{ Given the technical nature of our work, it is useful to comment and highlight the physical significance of our results. To this end, it is important to stress that all previous studies regarding the one-particle density matrix in the anyonic Lieb-Liniger model were restricted to infinite values of the repulsive interactions. This is because, although a Bethe Ansatz solution of the interacting Hamiltonian has been known for a long-time~\cite{Patu07,Batchelor2007}, only recently a formula for the field form factor has been derived~\cite{Piroli2020}. Accordingly, the main source of relevance of the present manuscript is to provide a set of predictions for the one-particle density matrix in the presence of interactions.
	
This is relevant, also in light of the most recent proposals for experimental implementations on $1D$ anyons: in particular, the possibility of tunable (finite) interactions  has been explicitly and successfully addressed in Refs.~\cite{gs-15,sse-16}, improving the scheme put forward in the seminal work~\cite{klmr-11}. We note that these papers propose the realization of an anyonic Hubbard chain. However, for low temperatures and fillings, the anyonic Lieb-Liniger model emerges as  a low-density limit of the latter, analogously to the well-known case of repulsive bosons in a lattice~\cite{Caza03,Caza04}. Thus, the explicit results reported in Secs.~\ref{sec:corr-homo} and \ref{sec:conf_pot} already give us realistic predictions for the one-particle density matrix in these experimental settings. Furthermore, our approach is flexible enough to be adjusted for more general trapping potentials that could appear in different setups.

The plots shown throughout the paper allow us to appreciate clear qualitative properties of the one-particle density matrix as a function of both the anyonic parameter $\kappa$ and the (tunable) interaction $c$. First of all, the one-particle density matrix $ g_1 (x, x') $ displays a strong dependence on the value of the parameter $\kappa$, irrespective of the trapping potential and interactions. In particular, greater values of $\kappa$ generically result in the presence of oscillations with increasing frequency. This feature is qualitatively similar to what has been observed in the infinitely-repulsive case~\cite{Santachiara2007,Calabrese2007,Santachiara2008,Marmorini16}, but it is enhanced away from the infinite-$c$ limit. This can be clearly seen in Fig.~\ref{fig:corr_pbc} for the case of periodic boundary conditions.  In the inhomogeneous case, we  also find a marked spatial dependence of correlations for finite $c$, which is much less pronounced in the limit of infinite interactions.}\\

{Finally, although our work has focused on the case of anyonic gases at equilibrium, it also provides the technical tools to move on to study non-equilibrium situations, where there are several interesting directions to be explored. For instance, while we know that many results for a gas expansion \cite{csc-13,rm-05,ck-12} easily generalise to anyons, less obvious is whether one could apply the recently developed techniques of Ref.~\cite{RCDD20}, where the authors extended the generalised hydrodynamics (GHD) formalism \cite{BCDF16,CaDY16} by taking into account quantum fluctuations.} This resulted in a {\it quantum GHD}, which, to some extent, can be viewed as a multi-component Luttinger liquid approach to the out-of-equilibrium dynamics in integrable systems. 
{ Our work provides the basis for  a generalization of this and other studies to the case of anyonic statistics, whose physics out-of-equilibrium has still received relatively small attention}.

\section*{Acknowledgments}
The authors are grateful to A Bastianello, Y Brun and J Dubail for very useful discussions and enlightening comments, especially regarding the numerical evaluation of Green's functions of Sec.~\ref{sec:corr-finite}. LP acknowledges support from the Alexander von Humboldt foundation. PC and SS acknowledge support from ERC under Consolidator grant  number 771536 (NEMO). 

\appendix

\section{Matching of the excited states}\label{app:matching}
In this appendix, we investigate the correspondence between the excited states of the anyonic Lieb-Liniger model \eqref{anyon-Lieb-Liniger} and the vertex operators \eqref{vertex} of the Luttinger liquid Hamiltonian \eqref{luttinger}. 
We recall that the eigenstates of \eqref{anyon-Lieb-Liniger} are determined by a set of rapidities $\{\lambda_j\}_{j=1}^N$ (see Eq.~\eqref{eigenstate}) that are labeled by the set of quantum numbers $I_j$ through the solution of Bethe equations \eqref{log-BAE}. In the ground state, the Bethe integers $I_j$ are uniformly distributed
\be\label{Bethe-int-GS}
\{I_j\}_{j=1}^N = \left\{j -\frac{1}{2}(N+1)\right\}_{j=1}^N.
\ee
Low-energy excitations of the model are of two main kinds (see e.g.~\cite{Giamarchi2003,Korepin1993}): ({\bf i}) change of the total particle number $N\to N\pm n$ and ({\bf ii}) particle-hole pair formation around the Fermi points $\pm Q$. Particle-hole excitations further divide into: ({\bf ii} {\it a}) particle-hole formation around the same Fermi point and ({\bf ii} {\it b}) backscattering processes where a particle (hole) is generated around one Fermi point $\pm Q$ and a hole (particle) is generated around the other $\mp Q$. 

For large systems $L\to \infty$, the contribution of low lying states to the energy and momentum eigenvalues can be estimated as \cite{Korepin1993}
\be\label{micro-E}
\delta E= \frac{2\pi v_s}{L}\left( \frac{n^2}{4 {\cal Z}^2} +\frac{m^2 \ {\cal Z}^2}{4} +\sum_{\ell=1}^\infty \ell(N_{\ell} +N_{-\ell})\right) +{\cal O}(L^{-2}),
\ee
\be\label{micro-P}
P= k_F m + \frac{2\pi}{L}\left(\frac{nm}{2} +\sum_{\ell=1}^\infty \ell(N_\ell-N_{-\ell})\right),
\ee
with Fermi momentum $k_F=\pi\rho$. Here, $n$ is the number of particles that have been added (removed) from the system ({\bf i}), $m$ is the number of particle-hole excitations across the Fermi sea ({\bf ii} {\it b}) and $N_{\pm\ell}$ is the number of particle-hole excitations around the same Fermi point ({\bf ii} {\it a}). The quantity ${\cal Z}$ is the dressed charge of the system given by ${\cal Z}=\sqrt{K}$ for Lieb-Liniger models \cite{Korepin1993}.

The Bethe integers configurations of such low-energy excitations can be obtained from \eqref{Bethe-int-GS} as follows. The change of the particle number ({\bf i}) corresponds to 
\be
\{I_j\}_{j=1}^N \to \left\{ j -\frac{1}{2}(N+1\pm n)\right\}_{j=1}^{N\pm n},
\ee
while a particle-hole formation ({\bf ii} {\it a}) with e.g. $N_{\pm 1}=1$ moves the max (min) Bethe integer of $\pm 1$
\be
N_1: \; \max(I_j)=\frac{N-1}{2}+1; \quad N_{-1}: \; \min(I_j)=-\frac{N-1}{2} -1,
\ee
and a backscattering process ({\bf ii} {\it b}) is a $\pm m$ shift of all the particles, entering in the Bethe Ansatz equations  \eqref{log-BAE} similarly to the anyonic $\kappa$ dependence.
An illustration of Bethe integers for different low-energy processes is given in Table~\ref{table:matching}.
\begin{table}[t]
\centering
\begin{tabular}{lc}
\toprule
Excited states &  Bethe integers \\
\midrule
Ground state & $\circ\circ\circ\bullet\bullet\bullet\bullet\bullet\bullet\bullet\bullet\circ\circ\circ$ \\
$n=1$ & $\circ\circ\circ\bullet\bullet\bullet\bullet \bullet \bullet\bullet\bullet\bullet\circ\circ\circ$\\
 $N_1=1$& $\circ\circ\circ\bullet\bullet\bullet\bullet\bullet\bullet\bullet\circ\bullet\circ\circ$ \\
$N_{-1}=1$ & $\circ\circ\bullet\circ\bullet\bullet\bullet\bullet\bullet\bullet\bullet\circ\circ\circ$ \\
 $N_2=1$ & $\circ\circ\circ\bullet\bullet\bullet\bullet\bullet\bullet\bullet\circ\circ\bullet\circ$ \\
 $N_1=2$ & $\circ\circ\circ\bullet\bullet\bullet\bullet\bullet\bullet\circ\bullet\bullet\circ\circ$ \\
$N_1=N_2=1$ & $\circ\circ\circ\bullet\bullet\bullet\bullet\bullet\bullet\circ\bullet\circ\bullet\circ$ \\
\bottomrule
\end{tabular}
\caption{Illustration of the Bethe {quantum numbers $I_j$} associated with the low-energy excitations of the model.
Here, the circles denotes a sequence of (half-)integers that are equally spaced. 
A full circle $\bullet$ stands for a {selected} value of $I_j$.}\label{table:matching}
\end{table}

On the other hand, we can estimate the energy and momentum contribution of the vertex operator $\A_{m,n}$ using  well-known conformal field theory results. At leading order, one obtains \cite{Korepin1993,Gogolin-Tsvelik}
\be\label{conf-E}
\delta E =\frac{2\pi v_s}{L} \Delta_{m,n},
\ee
and 
\be \label{conf-P}
 P = k_F m + \frac{2\pi}{L} \ s_{m,n},
\ee
where $\Delta_{m,n}$ is given in Eq.~\eqref{scaling} and $s_{m,n}= n m/2$ is the conformal spin associated with the vertex operator $\A_{m,n}$. 

The comparison of Eqs.~\eqref{conf-E}, \eqref{conf-P} with Eqs.~\eqref{micro-E}, \eqref{micro-P} allows us to match the excited state $\ket{\mu_j}_{j=1}^{N-1}$ (obtained from the Bethe integers configuration $\{I_j\}_{j=1}^{N-1}= \left\{ j -\frac{N}{2}  \right\}_{j=1}^{N-1}$ with $2m+\kappa$ shift of quasimomenta) and the vertex operator $\A_{2m+\kappa,-1}$.

Finally, particle-hole excitations of type ({\bf ii} {\it a}) are recovered in the Luttinger liquid theory as derivatives of the field $\hat\phi$.

\section{Non-universal amplitudes $B^\kappa_m$ from field form factors}\label{app:coefficient}
In this appendix, we provide a proof of the formula in Eq.~\eqref{coef-formula} that has been used in the main text to evaluate the non-universal amplitudes $B_m^\kappa$ appearing in the asymptotic expansion \eqref{bosonization} of $\Psikd$. The starting point is the asymptotic representation of a local field $\hat{O}_n(x)$ of our microscopic model \eqref{anyon-Lieb-Liniger} as a combination of operators $\hat\varphi_{m,n}(x)$ of the effective field theory \eqref{luttinger}
\be\label{exp}
\hat{O}_n(x)=\sum_m  c_m \ \hat\varphi_{m,n}(x).
\ee
The quantum number $n$ is a characteristic of the operator $\hat{O}_n$. For the anyonic Lieb-Liniger model \eqref{anyon-Lieb-Liniger}, it is related to the change in the total particles number $N$ generated by $\hat{O}_n$, see \ref{app:matching}.

We then need to specify the boundary conditions for the microscopic model and for its asymptotic description. Periodic boundary conditions for $\hat{O}_n$ are implemented in the r.h.s. of Eq.~\eqref{exp} as
\be\label{exp-pbc}
\hat{O}_n(x)=\sum_m \left(\frac{2\pi}{L}\right)^{\Delta_{m,n}}  c_m \ \hat\varphi_{m,n}(e^{\frac{2\pi i x}{L}}),
\ee
where $\Delta_{m,n}$ is the scaling dimension of $\hat\varphi_{m,n}$. In \ref{app:matching}, we have seen that, at low-energy and for large system sizes, there is an unambiguous matching between eigenstates $\ket{e_{m,n}}_L$ of the microscopic model and eigenstates $\ket{e_{m,n}}$ of the asymptotic field theory. In the limit $L\to\infty$, finite-size effects drop out and these eigenstates will eventually coincide. 
Therefore, if we {take the expectation}  $_L\braket{0| \cdot |e_{m,n}}_L$ of both sides of Eq.~\eqref{exp-pbc}, we have
\be
_L\braket{0| \hat{O}_n(x)|e_{m,n}}_L = \left(\frac{2\pi}{L}\right)^{\Delta_{m,n}}  c_m \times \  _L\braket{0|\hat\varphi_{m,n}(e^{\frac{2\pi i x}{L}})|e_{m,n}^\kappa}_L,
\ee
that, in the thermodynamic limit $L\to\infty$, leads to
\be\label{coef-d1}
c_m=\limTh\left[ \left(\frac{L}{2\pi}\right)^{\Delta_{m,n}}\  _L\braket{0| \hat{O}_n(x)|e_{m,n}}_L\right],
\ee
since $\braket{0|\hat\varphi^\kappa_{m,n}(0)|e_{m,n}}=\braket{e_{m,n}|e_{m,n}}=1$. 

The coefficient $c_m$ depends on the spatial position as \cite{Korepin1993}
\be\label{phase}
c_m= B_m \ e^{i k_F m x},
\ee
due to the momentum carried by the excited state $\ket{e_{m,n}}$, see Eqs.~\eqref{micro-P} and \eqref{conf-P} in \ref{app:matching}.  
Therefore, focusing on the amplitude $B_m$, one can write Eq.~\eqref{coef-d1} as
\be\label{coef-d2}
B_m=\limTh\left[ \left(\frac{L}{2\pi}\right)^{\Delta_{m,n}}\  _L\braket{0| \hat{O}_n(0)|e_{m,n}}_L\right].
\ee
Let us now consider  the case of interest, where the local field $\hat{O}_n=\Psikd$ with $n=-1$, since the operator $\Psikd$ is responsible for the creation of a particle and $\hat\varphi_{m,n}=\A_{m,n}$ are the vertex operators given in Eq.~\eqref{vertex}. From Eq.~\eqref{coef-d2} we have
\be\label{coef-last}
B_m^\kappa=\limTh\left[ \left(\frac{L}{2\pi}\right)^{\Delta_{2m+\kappa,1}}\  \frac{\braket{\{\lambda\}_{j=1}^N| \Psikd(0)|\{\mu_j\}_{j=1}^{N-1}}}{\sqrt{\braket{\{\lambda_j \}|\{\lambda_j\}}}\sqrt{\braket{\{\mu_j \}|\{\mu_j\}}}}\right],
\ee
where $\ket{0}_L=\ket{\{\lambda_j\}}$ is the $N$-particle ground state of the model \eqref{anyon-Lieb-Liniger} and $\ket{e_{2m+\kappa,-1}}_L=\ket{\{\mu_j\}}$ is the $(N-1)$-particle state associated with the vertex operator $\A_{2m+\kappa,-1}$, see \ref{app:matching}.

The result \eqref{coef-last} is the formula in Eq.~\eqref{coef-formula} that we used in the main text for the computation of non-universal amplitudes $B_m^\kappa$. Notice that the field form factor
\be
\mathbf{F}_{N, N-1}\left[x;\{\lambda_j\}_{j=1}^{N}, \{\mu_j\}_{j=1}^{N-1}\right]\equiv \braket{\{\lambda\}_{j=1}^N| \Psikd(x)|\{\mu_j\}_{j=1}^{N-1}}
\ee
satisfies \cite{Piroli2020}
\bea
\mathbf{F}_{N, N-1}\left[x;\{\lambda_j\}_{j=1}^{N}, \{\mu_j\}_{j=1}^{N-1}\right]=&\exp\left(i\left(P[\{\mu_j\}_{j=1}^{N-1}] -P[\{\lambda_j\}_{j=1}^{N}]\right) x\right)\nonumber\\[3pt]
& \times \mathbf{F}_{N, N-1}\left[0;\{\lambda_j\}_{j=1}^{N}, \{\mu_j\}_{j=1}^{N-1}\right],
\eea
consistently with Eq.~\eqref{coef-d1}-\eqref{coef-d2} in the limTh. With Eq.~\eqref{coef-last} at hand, we numerically evaluated the amplitude $B_m^\kappa$ for different values of $N$ ranging from $N=15$ to $N=30$, setting $L$ so that $N/L=\rho$ (hence $\gamma=c/\rho$) is kept fixed. Afterwards, we extrapolated the thermodynamic limit value with a polynomial fit in $1/N$. Note that the evaluation of the field form factor $\mathbf{F}_{N,N-1}$ for large values of $N$ requires the knowledge of the determinant formula derived in Ref.~\cite{Piroli2020}. Indeed, the numerical evaluation of the multidimensional integrals involved in $\mathbf{F}_{N,N-1}$ is already highly non-trivial for $N=5$ and it quickly becomes  impossible for higher $N$ due to the increasing complexity of the Bethe wavefunctions.

\section{Exact results for Green's functions with uniform $K$}\label{app:exact}
{It is useful to consider in more detail the analytical results for the Green's functions \eqref{Gpp}-\eqref{Gpt} in a uniform medium $K$.  Indeed, on the one hand these represent a (non-trivial) test for our implementation of the numerical algorithm in \ref{app:green}.
On the other hand, they clearly show how the short-distance divergence of the Green's functions \eqref{Gpp}-\eqref{Gtt} can be regularized. 
For uniform $K$, the Green's functions are} \cite{Cazalilla2004,Brun2017,Brun2018} 
\bea
\label{G-pp-ana}
\braket{\hat\phi(r)\ \hat\phi(r^\prime)}_{\mathrm{obc}}=\frac{-K}{4} \log \left(\left|\sin\left[\frac{\pi(r-r^\prime)}{2}\right]\right|^2  \left|\sin\left[\frac{\pi(r+r^\prime)}{2}\right]\right|^{-2} \right),\\
\label{G-tt-ana}
\braket{\hat\theta(r)\ \hat\theta(r^\prime)}_{\mathrm{obc}}=\frac{-1}{4K}  \log \left(\left|2\sin\left[\frac{\pi(r-r^\prime)}{2}\right]\right|^2  \left|2\sin\left[\frac{\pi(r+r^\prime)}{2}\right]\right|^{2} \right),\\
\label{G-pt-ana}
\braket{\hat\phi(r)\ \hat\theta(r^\prime)}_{\mathrm{obc}}=\frac{i\pi}{4}\sgn(r-r^\prime).
\eea
It is then easy to see that \eqref{G-pp-ana} and \eqref{G-tt-ana} diverge when $r\to r^\prime$ as
\be
\lim_{r\to r^\prime} G_{\phi\phi}(r,r^\prime)= -\frac{K}{4} \log|r-r^\prime|^2,
\ee
and the same for $G_{\theta\theta}$ with $K\leftrightarrow 1/K$. Therefore, if we remove such singular part for $r\to r^\prime$ from Eqs.~\eqref{G-pp-ana}-\eqref{G-tt-ana}, we obtain the regularised functions 
\bea\label{TG-reg1}
G^\reg_{\phi\phi}(r)=-\frac{K}{4}\log \left(\left|2\sin(\pi r)\right|^{-2}\right)  ,\\
\label{TG-reg2}
G^\reg_{\theta\theta}(r)=-\frac{1}{4K}\log \left(\left|2\sin(\pi r)\right|^{2}\right) .
\eea
Note that this prescription can be straightforwardly extended to inhomogeneous settings (because $K(r)$ is locally uniform when $r\to r^\prime$) and leads to the relations \eqref{Gpp-reg}-\eqref{Gtt-reg} of the main text.
It is also easy to check that plugging Eqs.~\eqref{G-pp-ana}-\eqref{G-pt-ana} and \eqref{TG-reg1}-\eqref{TG-reg2} in \eqref{inh_vertex} with $K=1$,  the analytical result of Eq.~\eqref{cazalilla} is recovered.

\section{Numerical computation of Green's functions}\label{app:green}
The numerical computation of the Green's functions \eqref{Gpp}-\eqref{Gpt} of the inhomogeneous Luttinger liquid Hamiltonian \eqref{luttinger} is made following the algorithm recently developed in Ref.~ \cite{Bastianello2020}. We start from the Hamiltonian \eqref{luttinger} (in normalised isothermal coordinates $r=\tilde{x}/\tilde{L}$)
\be
\hat{H}_{LL}=\frac{1}{2\pi}\int_0^1 \dd r \, \left(K(r)\ \pi^2\hat\Pi(r) + \frac{1}{K(r)} (\de_{r}\hat\phi)^2\right),
\ee
in terms of the field $\pi\hat\Pi(r)\equiv\de_r\hat\theta(r)$, $[\hat\Pi(r),\hat\phi(r^\prime)]=-i\delta(r-r^\prime)$ \cite{Giamarchi2003}, and we consider the following lattice discretisation 
\be\label{luttinger-discrete}
\hat{H}^{(\Lambda)}_{LL}=\frac{1}{2\pi \Lambda} \sum_{j=1}^{\Lambda} \pi^2 K_j \hat\Pi_j + \frac{\Lambda}{\pi} \sum_{j=1}^{\Lambda+1} \frac{1}{K_j+K_{j-1}} \left(\hat\phi_j-\hat\phi_{j-1}\right)^2,
\ee
where $\Lambda$ is the number of sampled points in the segment $[0,1]$ and $\hat\Pi_j$, $\hat\phi_j$ are the lattice discretisation of the fields $\hat\Pi(r)$, $\hat\phi(r)$ satisfying $[\hat\Pi_j,\hat\phi_{j^\prime}]=-i\delta_{j,j^\prime}$. The system is taken with open boundary conditions that imply $\hat\phi_0=\hat\phi_{\Lambda+1}=0$ and $K_0=K_{\Lambda+1}=1$. 

We then proceed with the implementation of the algorithm. For completeness, we report the main steps of the procedure, addressing the reader to Ref.~\cite{Bastianello2020} for further information. First, we introduce the ladder combinations
\be\label{ladder-comb}
\hat\varphi^+_j\equiv\frac{1}{\sqrt{2}}\left(\hat\phi_j + i \hat\Pi_j\right), \quad 
\hat{\varphi}^-_j\equiv\frac{1}{\sqrt{2}}\left(\hat\phi_j - i \hat\Pi_j\right),
\ee
satisfying bosonic commutation relations $[\hat\varphi^+_j,\hat\varphi^-_k]=\delta_{j,k}$. We subsequently cast $\hat\varphi_j^\pm$ inside the $2\Lambda$-vector
\be
\hat{\bm \varphi}^\dagger=\left[ \hat\varphi^-_1, \dots , \hat\varphi^-_{\Lambda}, \hat\varphi^+_1,\dots, \hat\varphi^+_{\Lambda}\right],
\ee
so that the lattice Hamiltonian \eqref{luttinger-discrete} is written as the quadratic form
\be\label{quadratic}
\hat{H}^{(\Lambda)}_{LL}= \hat{\bm \varphi}^{\dagger} 
\left[\begin{array}{cc}  A  & B \\  B^\dagger  &  A \end{array} \right]
\hat{\bm \varphi},
\ee
where $A, B$ are $\Lambda\times\Lambda$ matrices with elements
\be
\fl
A_{i,j}=\delta_{i,j} \left[ \frac{\pi K_j}{4\Lambda} + \frac{\Lambda}{2\pi (K_{j+1}+K_j)} + \frac{\Lambda}{2\pi (K_j +K_{j-1})}\right] -\frac{\Lambda \delta_{|i-j|,1}}{2\pi(K_j+K_i)},
\ee
and
\be
\fl
B_{i,j}=\delta_{i,j} \left[-\frac{\pi K_j}{4\Lambda} + \frac{\Lambda}{2\pi (K_{j+1}+K_j)} + \frac{\Lambda}{2\pi (K_j +K_{j-1})}\right] -\frac{\Lambda \delta_{|i-j|,1}}{2\pi(K_j+K_i)}.
\ee
Next, we consider the unitary transformation ${\cal U}$, ${\bm \Gamma}={\cal U}{\bm \varphi}$, that diagonalises the matrix in Eq.~\eqref{quadratic}
\be
\hat{H}^{(\Lambda)}_{LL}= \hat{\bm \Gamma}^{\ \dagger} {\cal U}^\dagger 
\left[\begin{array}{cc}  A  & B \\  B^\dagger  &  A \end{array} \right] 
{\cal U} \hat{\bm \Gamma}= \sum_{j=1}^{2\Lambda} \lambda_j \hat{\bm \eta}^\dagger_j \hat{\bm \eta}_j,
\ee
where the eigenvectors have the structure
\be
\hat{\bm \eta}^\dagger=\left[\hat\eta^\dagger_1, \dots, \hat\eta^\dagger_{\Lambda}, \hat\eta_1,\dots, \hat\eta_\Lambda\right],
\ee
which follows from the structure of the Fock space. In order to preserve the bosonic commutation relations of fields $\hat\varphi^\pm_j$, the matrix ${\cal U}$ must have a simplectic structure
\be
{\cal U} \left[\begin{array}{cc} \mathbb{I} & 0 \\ 0 & -\mathbb{I} \end{array}\right] {\cal U}^\dagger =\left[\begin{array}{cc} \mathbb{I} & 0 \\ 0 & -\mathbb{I} \end{array}\right],
\ee
leading to 
\be
\hat{H}^{(\Lambda)}_{LL}=  \sum_{j=1}^{\Lambda} \lambda_j \left( \hat\eta_j^\dagger\hat\eta_j + \hat\eta_j\hat\eta^\dagger_j\right),
\ee
with $\lambda_j> 0$. The ground state correlation matrix of the model \eqref{luttinger-discrete} is written as the $2\Lambda\times 2\Lambda$ hermitian matrix
\be
\braket{ \hat{\bm\varphi} \hat{\bm \varphi}^\dagger}={\cal U} \braket{\hat{\bm\Gamma} \hat{\bm\Gamma}^\dagger} {\cal U}^\dagger,
\ee
that can be expressed as the projector over positive frequencies
\be\label{corr-matrix}
\braket{ \hat{\bm\varphi} \hat{\bm \varphi}^\dagger}=\sum_{\omega_j >0} {\bm v}_j {\bm v}^\dagger_j,
\ee 
where ${\bm v}_j$ are eigenstates of the matrix
\be
 {\cal H}=\left[\begin{array}{cc} \mathbb{I} & 0 \\ 0 & -\mathbb{I} \end{array}\right] \left[\begin{array}{cc}  A  & B \\  B^\dagger  &  A \end{array}\right].
\ee
Clearly, the eigenproblem  ${\cal H}{\bm v}_j = \omega_j {\bm v}_j$ is not enough to properly fix the orthonormality conditions on the vectors ${\bm v}_j$. This can be done as follows. We consider a set of eigenvectors $\{{\bm w}_j\}$ that spans the positive spectrum of ${\cal H}$. Next, we construct the hermitian matrix
\be
O_{i,j}= {\bm w}_i^\dagger \left[\begin{array}{cc} \mathbb{I} & 0 \\ 0 & -\mathbb{I} \end{array} \right]{\bm w}_j,
\ee
and we consider the unitary transformation ${\cal B}$ such that ${\cal B}^\dagger O {\cal B}$ is diagonal. An orthogonal set of vectors is then build as
\be
\tilde{\bm w}_j =\sum_i {\bm w}_i \ {\cal B}_{i,j},
\ee
and finally normalised vectors appearing in Eq.~\eqref{corr-matrix} are obtained as
\be
{\bm v}_j = \left| \tilde{\bm w}_j^\dagger  \left[\begin{array}{cc} \mathbb{I} & 0 \\ 0 & -\mathbb{I} \end{array}\right] \tilde{\bm w}_j\right|^{-1/2}  \tilde{\bm w}_j.
\ee

From the correlation matrix $C\equiv\braket{ \hat{\bm\varphi} \hat{\bm \varphi}^\dagger}$ in Eq.~\eqref{corr-matrix}, it is easy to extract the two-point correlators of fields $\hat\phi_j$ and $\hat\Pi_j$ using  Eq.~\eqref{ladder-comb}. For instance, $\braket{\hat\phi_j\hat\phi_k}$ is given by
\be
\braket{\hat\phi_j\hat\phi_k}=\frac{1}{2\Lambda}\left( C_{j,k} + C_{j+\Lambda,k} + C_{j,\Lambda+k} + C_{j+\Lambda,k+\Lambda}  \right),
\ee
and similarly for the others. As final result, from the algorithm of Ref.~\cite{Bastianello2020}, we obtain the $2\Lambda\times 2\Lambda$ correlation matrix
\be\label{corr_matrix}
C_{\phi\Pi}=\begin{blockarray}{[cc]}
\left[ \braket{\hat\phi_j\hat\phi_k}\right]_{j,k=1}^\Lambda  &\left[\braket{\hat\Pi_j\hat\phi_k} \right]_{j,k=1}^\Lambda  \\[10pt]
\left[\braket{\hat\phi_j\hat\Pi_k} \right]_{j,k=1}^\Lambda  & \left[ \braket{\hat\Pi_j\hat\Pi_k} \right]_{j,k=1}^\Lambda 
\end{blockarray}\ .
\ee
\begin{figure}[t]
\centering
\includegraphics[width=0.6\textwidth]{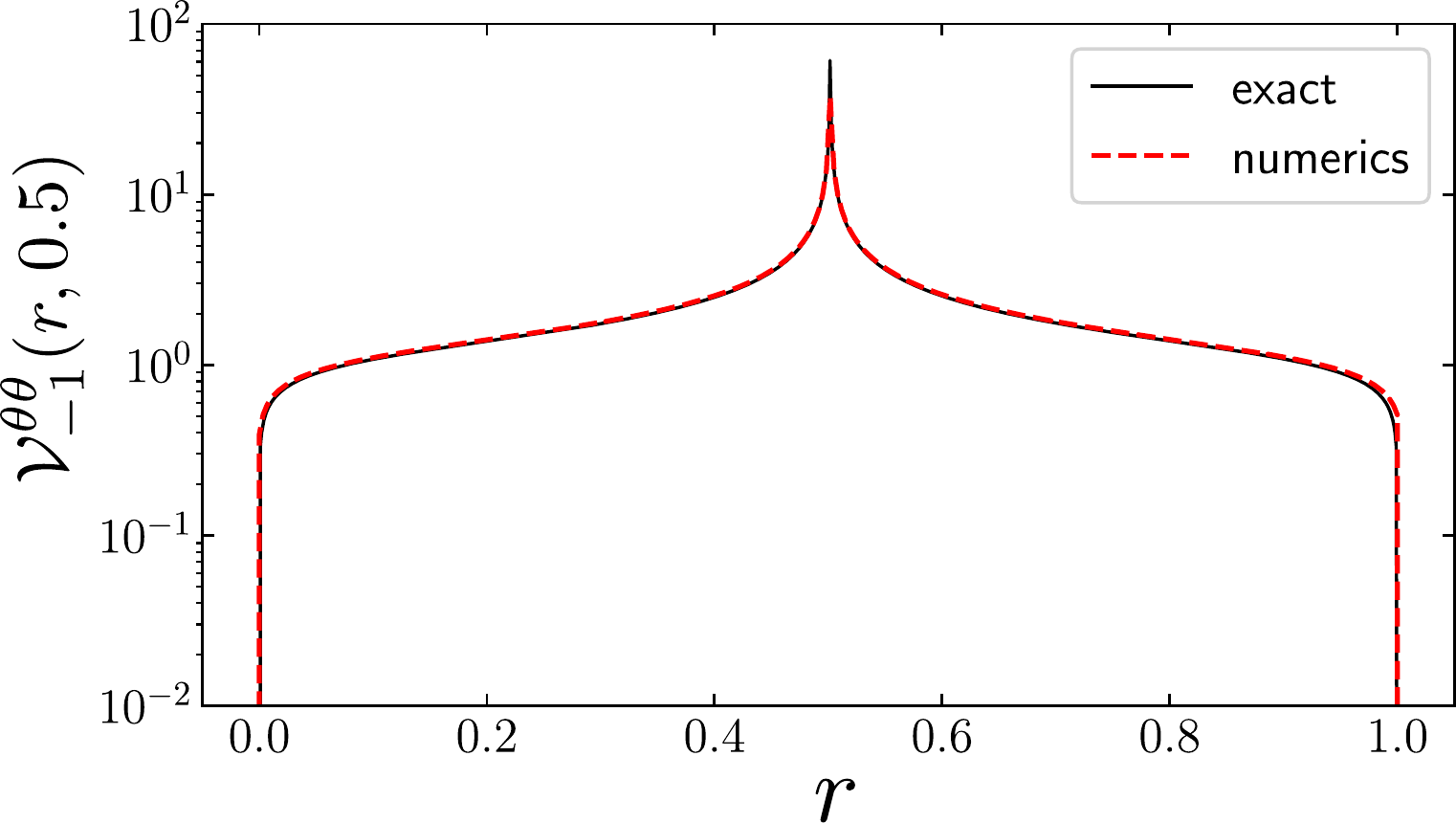}
\caption{Comparison between the numerically computed quantity ${\cal V}^{\theta\theta}_{-1}$ in \eqref{match-vertex} and its exact value for $K=1$ (see Eqs.~\eqref{G-pp-ana}-\eqref{TG-reg2} and \eqref{Vtt}). Data are obtained with the algorithm of Ref.~\cite{Bastianello2020}. An excellent agreement of the curves is already found for a modest sampling $\Lambda=300$.}\label{fig:match-vertex}
\end{figure}

At this point, we are left with the numerical integration of the $\phi$-$\Pi$ correlation matrix \eqref{corr_matrix} in order to get the $\phi$-$\theta$ correlation matrix we are interested in. Note that the integration of the field $\hat\Pi$
\be
\hat\theta(r)=\hat\theta(0)+\pi\int_0^{r} \dd y \ \hat\Pi(y),
\ee
gives rise to a global phase $\hat\theta(0)$ that we are unable to determine. However, the one-particle density matrix only requires the knowledge of the quantity (see Eq.~\eqref{inh_vertex} and \eqref{Vtt}-\eqref{1pdm_final})
\be\label{match-vertex}
{\cal V}^{\theta\theta}_{-1}(r,r^\prime)\equiv\braket{e^{-i\hat\theta(r)} \ e^{i\hat\theta(r^\prime)}},
\ee
which is, by construction, independent on $\hat\theta(0)$. The comparison between the numerically calculated quantity ${\cal V}^{\theta\theta}_{-1}$ in \eqref{match-vertex} and the exact result for $K=1$ (see Eqs.~\eqref{G-pp-ana}-\eqref{TG-reg2} in \ref{app:exact}) gives an excellent agreement, as shown in Fig.~\ref{fig:match-vertex}. Similar arguments apply for the mixed Green's function $G_{\phi\theta}$.
\begin{figure}[t]
\centering
\includegraphics[width=\textwidth]{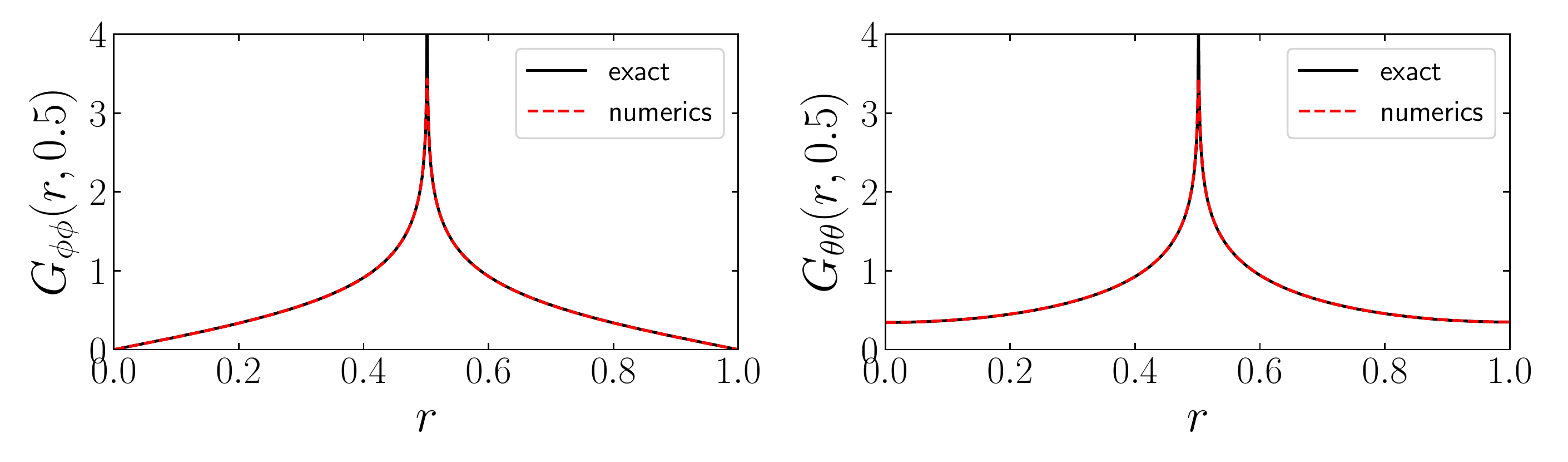}
\caption{Numerical evaluation of the Green's functions $G_{\phi\phi}$ ({\it left}) and $G_{\theta\theta}$ ({\it right}) compared with their analytical results in \eqref{G-pp-ana}, \eqref{G-tt-ana} for $K=1$. The agreement between the curves is excellent. Data are obtained with a numerical inversion of the generalised Laplace operator, see Eqs.~\eqref{GD}-\eqref{GN}.}
\label{fig:check}
\end{figure}
Finally, we mention that it is possible to numerically evaluate the Green's function $G_{\theta\theta}$ and $G_{\phi\phi}$ with other strategies, e.g. exploiting the duality of the gaussian free field theory of Eq.~\eqref{action-isothermal} \cite{Brun2018}. Indeed, by definition,  the Green's function $G_{\phi\phi}$ satisfies  (recall $\x=(\tilde{x},\tau)\in \Omega$)
\be\label{GD}
\nabla_\x \frac{1}{K(\x)} \nabla_\x \ G_{\phi\phi}(\x,\x^\prime)=4\pi \delta^{(2)}(\x-\x^\prime),
\ee
with pbc on the coordinate $\tau$ and open boundary conditions (Dirichlet conditions)
\be
G_{\phi\phi}(\x,\x^\prime)=0 \quad {\rm for}\quad  \x\in\de\Omega.
\ee
On the other hand, the Green's function  $G_{\theta\theta}$ is obtained replacing $K(\x) \leftrightarrow 1/K(\x)$
\be\label{GN}
\nabla_\x K(\x) \nabla_\x \ G_{\theta\theta}(\x,\x^\prime)=4\pi \delta^{(2)}(\x-\x^\prime)-4\pi/{\rm Vol},
\ee
and Dirichlet boundary conditions with Neumann boundary conditions (that are dual to Dirchlet ones)
\be
\nabla_\x G_{\theta\theta}(\x,\x^\prime)=0 \quad {\rm for}\quad  \x\in\de\Omega.
\ee
The second term on the r.h.s. of Eq.~\eqref{GN} is needed to annihilate unwanted zero modes of the Green's function $G_{\theta\theta}$ \cite{Brun2018}. In practice, one can numerically implement the generalised Laplace operators that multiply the Green's functions on the l.h.s. of Eqs.~\eqref{GD}-\eqref{GN} as matrices acting on a discretised two-dimensional space (with the correct type of boundary conditions) and derive the correlators with a numerical inversion of the kernel. In Fig.~\ref{fig:check}, the Green's functions obtained numerically with the kernel-inversion method are compared with the exact results for uniform $K$ (see Eqs.~\eqref{G-pp-ana}-\eqref{G-tt-ana} in \ref{app:exact}), showing a perfect agreement.




\addcontentsline{toc}{section}{References}


\Bibliography{100}



\bibitem{lm_77}
J. M. Leinaas and J. Myrheim, 
\href{http://dx.doi.org/10.1007/BF02727953}{Nuovo Cimento B {\bf 37}, 1 (1977)};\\
F. Wilczek, 
\href{http://dx.doi.org/10.1103/PhysRevLett.48.1144}{Phys. Rev. Lett. {\bf 48}, 1144 (1982)};\\
F. Wilczek, 
\href{http://dx.doi.org/10.1103/PhysRevLett.49.957}{Phys. Rev. Lett. {\bf 49}, 957 (1982)}.

\bibitem{laughlin_83}
R. B. Laughlin, 
\href{http://dx.doi.org/10.1103/PhysRevLett.50.1395}{Phys. Rev. Lett. {\bf 50}, 1395 (1983)};\\
B. I. Halperin, 
\href{http://dx.doi.org/10.1103/PhysRevLett.52.1583}{Phys. Rev. Lett. {\bf 52}, 1583 (1984)};\\
F. Wilczek, Fractional Statistics and Anyon Superconductivity, (World Scientific, Singapore 1990);\\
F. E. Camino, W. Zhou, and V. J. Goldman, 
\href{http://dx.doi.org/10.1103/PhysRevB.72.075342}{Phys. Rev. B {\bf 72}, 75342 (2005)};\\
E.-A. Kim, M. Lawler, S. Vishveshwara, and E. Fradkin, 
\href{http://dx.doi.org/10.1103/PhysRevLett.95.176402}{Phys. Rev. Lett. {\bf 95}, 176402 (2005)}.


\bibitem{klmr-11}
T. Keilmann, S. Lanzmich, I. McCulloch, and M. Roncaglia, 
\href{http://dx.doi.org/10.1038/ncomms1353}{Nature Comm. {\bf 2}, 361 (2011)}.

\bibitem{gs-15}
S. Greschner and L. Santos, 
\href{http://dx.doi.org/10.1103/PhysRevLett.115.053002}{Phys. Rev. Lett. {\bf 115}, 53002 (2015)}.

\bibitem{sse-16}
C. Str\"{a}ter, S. C. L. Srivastava, and A. Eckardt, 
\href{http://dx.doi.org/10.1103/PhysRevLett.117.205303}{Phys. Rev. Lett. {\bf 117}, 205303 (2016)}.

\bibitem{gcs-18}
S. Greschner, L. Cardarelli, and L. Santos, 
\href{https://doi.org/10.1103/PhysRevA.97.053605}{Phys. Rev. A {\bf 97}, 053605  (2018)}.


\bibitem{agjp-96}
U. Aglietti, L. Griguolo, R. Jackiw, S.-Y. Pi, and D. Seminara, 
\href{http://dx.doi.org/10.1103/PhysRevLett.73.2150}{Phys. Rev. Lett. {\bf 77}, 4406 (1996)}.

\bibitem{rabello-96}
S. J. Benetton Rabello, 
\href{http://dx.doi.org/10.1103/PhysRevLett.76.4007}{Phys. Rev. Lett. {\bf 76}, 4007 (1996)}.

\bibitem{it-99}
N. Ilieva and W. Thirring, 
\href{http://dx.doi.org/10.1007/BF02557229}{Theor. Math. Phys. {\bf 121}, 1294 (1999)};\\
N. Ilieva and W. Thirring, 
\href{http://dx.doi.org/10.1088/1751-8113/40/50/004}{Eur. Phys. J. C {\bf 6}, 705 (1999)}.

\bibitem{Kundu99}
A. Kundu, 
\href{http://dx.doi.org/10.1103/PhysRevLett.83.1275}{Phys. Rev. Lett. {\bf 83}, 1275 (1999)}.

\bibitem{lmp-00}
A. Liguori, M. Mintchev, and L. Pilo, 
\href{http://dx.doi.org/10.1016/S0550-3213(99)00774-9}{Nucl. Phys. B {\bf 569}, 577 (2000)}.

\bibitem{Girardeau06}
M. D. Girardeau, 
\href{http://dx.doi.org/10.1103/PhysRevLett.97.100402}{Phys. Rev. Lett. {\bf 97}, 100402 (2006)}.

\bibitem{an-07}
D. V. Averin and J. A. Nesteroff, 
\href{http://dx.doi.org/10.1103/PhysRevLett.99.096801}{Phys. Rev. Lett. {\bf 99}, 96801 (2007)}.

\bibitem{Batchelor06}
M. T. Batchelor, X.-W. Guan, and N. Oelkers, 
\href{http://dx.doi.org/10.1103/PhysRevLett.96.210402}{Phys. Rev. Lett. {\bf 96}, 210402 (2006)};\\
M. T. Batchelor and X.-W. Guan, 
\href{http://dx.doi.org/10.1103/PhysRevB.74.195121}{Phys. Rev. B {\bf 74}, 195121 (2006)}.

\bibitem{Patu07}
O. I. P\^{a}tu, V. E. Korepin, and D. V. Averin, 
\href{http://dx.doi.org/10.1088/1751-8113/40/50/004}{J. Phys. A: Math. Theor. {\bf 40}, 14963 (2007)}.

\bibitem{Batchelor2007}
M. T. Batchelor, X.-W. Guan, and J.-S. He, 
\href{http://dx.doi.org/10.1088/1742-5468/2007/03/P03007}{J. Stat. Mech. P03007  (2007) }.


\bibitem{bfglz-08}
M. T. Batchelor, A. Foerster, X.-W. Guan, J. Links, and H.-Q. Zhou, 
\href{https://doi.org/10.1088/1751-8113/41/46/465201}{J. Phys. A {\bf 41}, 465201 (2008)}.

\bibitem{BeCM09} 
B. Bellazzini, P. Calabrese, and M. Mintchev, 
\href{http://dx.doi.org/10.1103/PhysRevB.79.085122}{Phys. Rev. B {\bf 79}, 085122 (2009)}.

\bibitem{yao-12}
Y.-L. Yao, J.-P. Cao, G.-L. Li, and H. Fan,
\href{https://doi.org/10.1088/1751-8113/45/4/045207}{J. Phys. A {\bf  45}, 045207 (2012)}.

\bibitem{spk-12}
R. A. Santos, F. N. C. Paraan, and V. E. Korepin,  \href{http://dx.doi.org/10.1103/PhysRevB.86.045123}{Phys. Rev. B {\bf 86}, 045123 (2012)}.



\bibitem{zgf-17}
W. Zhang, S. Greschner, E. Fan, T. C. Scott, and Y. Zhang, 
\href{https://doi.org/10.1103/PhysRevA.95.053614}{Phys. Rev. A {\bf 95}, 053614 (2017)}.

\bibitem{yp-17}
L. Yang and H. Pu, 
\href{https://doi.org/10.1103/PhysRevA.95.051602}{Phys. Rev. A {\bf 95}, 051602 (2017)}.

\bibitem{jy-18}
H. H. Jen and S.-K. Yip, 
\href{https://doi.org/10.1103/PhysRevA.98.013623}{Phys. Rev. A {\bf 98}, 013623 (2018)}.

\bibitem{rccm-19}
D. Rossini, M. Carrega, M. Calvanese Strinati, and L. Mazza, 
\href{https://doi.org/10.1103/PhysRevB.99.085113}{Phys. Rev. B {\bf 99}, 085113 (2019)}.

\bibitem{mwd-20}
I. Mahyaeh, J. Wouters, and D. Schuricht, 
\href{https://arxiv.org/abs/2003.07812}{ArXiv:2003.07812 (2020)}.

\bibitem{CSM1}
F. Calogero, 
\href{ https://doi.org/10.1063/1.1664821 }{J.  Math. Phys. {\bf 10}, 2197 (1969)}\\
F. Calogero,
\href{https://doi.org/10.1063/1.1665604 }{J. Math. Phys. {\bf 12}, 419 (1971)}

\bibitem{CSM2}
B. Sutherland, 
\href{https://doi.org/10.1063/1.1665584}{J. Math. Phys. {\bf 12} (1971) 246}\\
B. Sutherland, 
\href{https://doi.org/10.1063/1.1665585}{J. Math. Phys. {\bf 12}(1971) 251}\\
B. Sutherland, 
\href{https://doi.org/10.1103/PhysRevA.4.2019}{Phys. Rev. A {\bf 4} (1971)} \\
B. Sutherland, 
\href{https://doi.org/10.1103/PhysRevA.5.1372}{Phys. Rev. A {\bf 5}(1971) 1372}

\bibitem{Lieb63} E. Lieb and W. Liniger, 
\href{http://dx.doi.org/10.1103/PhysRev.130.1605}{Phys. Rev. {\bf 130}, 1605 (1963)}; \\
E. Lieb, 
\href{http://dx.doi.org/10.1103/PhysRev.130.1616}{Phys. Rev. {\bf 130}, 1616 (1963)}.


\bibitem{Korepin1993} V.E. Korepin, N.M. Bogoliubov and A.G. Izergin, 
{\it Quantum inverse scattering method and correlation functions}, Cambridge University Press (1993). 


\bibitem{JiMi81} 
M. Jimbo and T. Miwa, 
\href{http://dx.doi.org/10.1103/PhysRevD.24.3169}{Phys. Rev. D {\bf 24}, 3169 (1981)}.

\bibitem{OlDu03} 
M. Olshanii and V. Dunjko, 
\href{http://dx.doi.org/10.1103/PhysRevLett.91.090401}{Phys. Rev. Lett. {\bf 91}, 090401 (2003)}.

\bibitem{GaSh03} 
D. M. Gangardt and G. V. Shlyapnikov, 
\href{http://dx.doi.org/10.1103/PhysRevLett.90.010401}{Phys. Rev. Lett. {\bf 90}, 010401 (2003)};\\
D. M. Gangardt and G. V. Shlyapnikov, 
\href{http://dx.doi.org/10.1088/1367-2630/5/1/379}{New J. Phys. {\bf 5}, 79 (2003)}.

\bibitem{ChSZ05} 
V. V. Cheianov, H. Smith, and M. B. Zvonarev, 
\href{http://dx.doi.org/10.1103/PhysRevA.71.033610}{Phys. Rev. A {\bf 71}, 033610 (2005)}.

\bibitem{CaCa06} 
J.-S. Caux and P. Calabrese, 
\href{http://dx.doi.org/10.1103/PhysRevA.74.031605}{Phys. Rev. A {\bf 74}, 31605 (2006)};\\
P. Calabrese and J.-S. Caux, 
\href{http://dx.doi.org/10.1103/PhysRevLett.98.150403}{Phys. Rev. Lett. {\bf 98}, 150403 (2007)}.

\bibitem{ChSZ06} 
V. V. Cheianov, H. Smith, and M. B. Zvonarev, 
\href{http://dx.doi.org/10.1103/PhysRevA.73.051604}{Phys. Rev. A {\bf 73}, 051604 (2006)};\\
V. V. Cheianov, H. Smith, and M. B. Zvonarev, 
\href{http://dx.doi.org/10.1088/1742-5468/2006/08/P08015}{J. Stat. Mech. P08015 (2006)}.

\bibitem{ScFl07} 
B. Schmidt and M. Fleischhauer, 
\href{http://dx.doi.org/10.1103/PhysRevA.75.021601}{Phys. Rev. A {\bf 75}, 021601 (2007)}.

\bibitem{kt-11}
K. K. Kozlowski and V. Terras, 
\href{http://dx.doi.org/10.1088/1742-5468/2011/09/P09013}{J. Stat. Mech. P09013 (2011)}.



\bibitem{kkmt-13}
N. Kitanine, K. K. Kozlowski, J. M. Maillet, and V. Terras, 
\href{http://dx.doi.org/10.1088/1742-5468/2014/05/P05011}{J. Stat. Mech. P05011 (2014)}

\bibitem{km-15}
K. K. Kozlowski and J. -M. Maillet, 
\href{http://dx.doi.org/10.1088/1751-8113/48/48/484004}{ J. Phys. A {\bf 48}, 484004 (2015)}


\bibitem{PiCa16-1} 
L. Piroli and P. Calabrese, 
\href{http://dx.doi.org/10.1103/PhysRevA.94.053620}{Phys. Rev. A {\bf 94}, 053620 (2016)}.


\bibitem{MiVT02} 
A. Minguzzi, P. Vignolo, and M. P. Tosi, 
\href{http://dx.doi.org/10.1016/S0375-9601(02)00042-7}{Phys. Lett. A {\bf 294}, 222 (2002)}.

\bibitem{KGDS03} 
K. V. Kheruntsyan, D. M. Gangardt, P. D. Drummond, and G. V. Shlyapnikov, 
\href{http://dx.doi.org/10.1103/PhysRevLett.91.040403}{Phys. Rev. Lett. {\bf 91}, 040403 (2003)}.

\bibitem{SGDV08} 
A. G. Sykes, D. M. Gangardt, M. J. Davis, K. Viering, M. G. Raizen, and K. V. Kheruntsyan, 
\href{http://dx.doi.org/10.1103/PhysRevLett.100.160406}{Phys. Rev. Lett. {\bf 100}, 160406 (2008)}.

\bibitem{kms-11}
K. K. Kozlowski, J. M. Maillet, and N. A. Slavnov, 
\href{http://dx.doi.org/10.1088/1742-5468/2011/03/P03018}{J. Stat. Mech. P03018 (2011)}.

\bibitem{kms-11b}
K. K. Kozlowski, J. M. Maillet, and N. A. Slavnov, 
 \href{http://dx.doi.org/10.1088/1742-5468/2011/03/P03019}{J. Stat. Mech. P03019 (2011)}.

\bibitem{KuLa13} 
M. Kulkarni and A. Lamacraft, 
\href{http://dx.doi.org/10.1103/PhysRevA.88.021603}{Phys. Rev. A {\bf 88}, 021603 (2013)}.

\bibitem{PaKl13} 
O. I. P\^{a}\c{t}u and A. Kl\"{u}mper, 
\href{http://dx.doi.org/10.1103/PhysRevA.88.033623}{Phys. Rev. A {\bf 88}, 033623 (2013)};\\
A. Kl\"{u}mper and O. I. P\^{a}\c{t}u, 
\href{http://dx.doi.org/10.1103/PhysRevA.90.053626}{Phys. Rev. A {\bf 90}, 053626 (2014)}.

\bibitem{ViMi13} 
P. Vignolo and A. Minguzzi, 
\href{http://dx.doi.org/10.1103/PhysRevLett.110.020403}{Phys. Rev. Lett. {\bf 110}, 020403 (2013)};\\
G. Lang, P. Vignolo, and A. Minguzzi, 
\href{http://dx.doi.org/10.1140/epjst/e2016-60343-6}{Eur. Phys. J. Spec. Top. {\bf 226}, 1583 (2017)}.

\bibitem{PaCa14} 
M. Panfil and J.-S. Caux, 
\href{http://dx.doi.org/10.1103/PhysRevA.89.033605}{Phys. Rev. A {\bf 89}, 033605 (2014)}.

\bibitem{NRTG16} 
E. Nandani, R. A. R\"{o}mer, S. Tan, and X.-W. Guan, 
\href{http://dx.doi.org/10.1088/1367-2630/18/5/055014}{New J. Phys. {\bf 18}, 055014 (2016)}.



\bibitem{KoMT09} 
M. Kormos, G. Mussardo, and A. Trombettoni, 
\href{http://dx.doi.org/10.1103/PhysRevLett.103.210404}{Phys. Rev. Lett. {\bf 103}, 210404 (2009)};\\
M. Kormos, G. Mussardo, and A. Trombettoni, 
\href{http://dx.doi.org/10.1103/PhysRevA.81.043606}{Phys. Rev. A {\bf 81}, 043606 (2010)}.

\bibitem{KoCI11} 
M. Kormos, Y.-Z. Chou, and A. Imambekov, 
\href{http://dx.doi.org/10.1103/PhysRevLett.107.230405}{Phys. Rev. Lett. {\bf 107}, 230405 (2011)}.

\bibitem{KoMT11_sTG} 
M. Kormos, G. Mussardo, and A. Trombettoni, 
\href{http://dx.doi.org/10.1103/PhysRevA.83.013617}{Phys. Rev. A {\bf 83}, 013617 (2011)}.

\bibitem{Pozs11_mv} 
B. Pozsgay, 
\href{http://dx.doi.org/10.1088/1742-5468/2011/01/P01011}{J. Stat. Mech. P01011 (2011)}.

\bibitem{Pozs11} 
B. Pozsgay, 
\href{http://dx.doi.org/10.1088/1742-5468/2011/11/P11017}{J. Stat. Mech. P11017 (2011)}.

\bibitem{PiCE16} 
L. Piroli, P. Calabrese, and F. H. L. Essler, 
\href{http://dx.doi.org/10.21468/SciPostPhys.1.1.001}{SciPost Physics {\bf 1}, 001 (2016)};\\
L. Piroli, P. Calabrese, and F. H. L. Essler, 
\href{http://dx.doi.org/10.1103/PhysRevLett.116.070408}{Phys. Rev. Lett. {\bf 116}, 070408 (2016)}.

\bibitem{BaPi18} 
A. Bastianello and L. Piroli, 
\href{http://dx.doi.org/10.1088/1742-5468/aaeb48}{J. Stat. Mech. 113104 (2018)}.

\bibitem{BaPC18} 
A. Bastianello, L. Piroli, and P. Calabrese, 
\href{http://dx.doi.org/10.1103/PhysRevLett.120.190601}{Phys. Rev. Lett. {\bf 120}, 190601 (2018)}.


\bibitem{CaCaSl07}
J.-S. Caux, P. Calabrese, and N. A. Slavnov, 
\href{http://dx.doi.org/10.1088/1742-5468/2007/01/P01008}{J. Stat. Mech.  P01008 (2007)}.

\bibitem{KoMP10} 
M. Kormos, G. Mussardo, and B. Pozsgay, 
\href{http://dx.doi.org/10.1088/1742-5468/2010/05/P05014}{J. Stat. Mech.  P05014 (2010)}.

\bibitem{PiCa15} 
L. Piroli and P. Calabrese, 
\href{http://dx.doi.org/10.1088/1751-8113/48/45/454002}{J. Phys. A: Math. Theor. {\bf 48}, 454002 (2015)}.



\bibitem{Patu08} 
O. I. P\^atu, V. E. Korepin, and D. V. Averin, 
\href{http://dx.doi.org/10.1088/1751-8113/41/14/145006}{J. Phys. A: Math. Theor. {\bf 41}, 145006 (2008)};\\
O. I. P\^atu, V. E. Korepin, and D. V. Averin, 
\href{http://dx.doi.org/10.1088/1751-8113/41/25/255205}{J. Phys. A: Math. Theor. {\bf 41}, 255205 (2008)}.


\bibitem{Patu09}
O. I. P\^atu, V. E. Korepin, and D. V. Averin, 
\href{http://dx.doi.org/10.1088/1751-8113/42/27/275207}{J. Phys. A: Math. Theor. {\bf 42}, 275207 (2009)};\\
O. I. P\^atu, V. E. Korepin, and D. V. Averin, 
\href{http://dx.doi.org/10.1088/1751-8113/43/11/115204}{J. Phys. A: Math. Theor. {\bf 43}, 115204 (2010)}.

\bibitem{Calabrese2007}
P. Calabrese and M. Mintchev, 
\href{http://dx.doi.org/10.1103/PhysRevB.75.233104}{Phys. Rev. B {\bf 75}, 233104 (2007)}.

\bibitem{Calabrese2009}
P. Calabrese and R. Santachiara, 
\href{http://dx.doi.org/10.1088/1742-5468/2009/03/P03002}{J. Stat. Mech.  P03002 (2009)}.

\bibitem{Patu19} 
O. I. P\^a\c{t}u, 
\href{http://dx.doi.org/10.1103/PhysRevA.100.063635}{Phys. Rev. A {\bf 100}, 063635 (2019)}.


\bibitem{Santachiara2007}
R. Santachiara, F. Stauffer, and D. C. Cabra, 
\href{http://dx.doi.org/10.1088/1742-5468/2007/05/L05003}{J. Stat. Mech. L05003 (2007)}.

\bibitem{ghc-09}
H. Guo, Y. Hao, and S. Chen, 
\href{http://dx.doi.org/10.1103/PhysRevA.80.052332}{Phys. Rev. A {\bf 80}, 52332 (2009)}.

\bibitem{Santachiara2008}
R. Santachiara and P. Calabrese, 
\href{http://dx.doi.org/10.1088/1742-5468/2008/06/P06005}{J. Stat. Mech.  P06005 (2008)}.

\bibitem{hzc-08}
Y. Hao, Y. Zhang, and S. Chen, 
\href{http://dx.doi.org/10.1103/PhysRevA.78.023631}{Phys. Rev. A {\bf 78}, 23631 (2008)}.

\bibitem{Patu09b}
O. I. P\^atu, V. E. Korepin, and D. V. Averin,
\href{http://dx.doi.org/10.1209/0295-5075/86/40001}{EPL {\bf 86}, 40001 (2009)}.

\bibitem{Patu15}
O. I. P\^atu, 
\href{http://dx.doi.org/10.1088/1742-5468/2015/01/P01004}{J. Stat. Mech.  P01004 (2015)}.

\bibitem{tep-15}
G. Tang, S. Eggert, and A. Pelster, 
\href{https://doi.org/10.1088/1367-2630/17/12/123016}{New J. Phys. {\bf 17}, 123016 (2015)}.

\bibitem{Zinn15} 
N. T. Zinner, 
\href{http://dx.doi.org/10.1103/PhysRevA.92.063634}{Phys. Rev. A {\bf 92}, 63634 (2015)}.

\bibitem{hao-16}
Y. Hao, 
\href{http://dx.doi.org/10.1103/PhysRevA.93.063627}{Phys. Rev. A {\bf 93}, 63627 (2016)}.

\bibitem{Marmorini16}
G. Marmorini, M. Pepe, and P. Calabrese, 
\href{http://dx.doi.org/10.1088/1742-5468/2016/07/073106}{J. Stat. Mech. 73106 (2016)}.

\bibitem{hs-17}
Y. Hao and Y. Song, 
\href{https://doi.org/10.1140/epjd/e2017-70501-8}{Eur. Phys. J. D {\bf 71}, 135 (2017)}.


\bibitem{del_Campo08}
A. del Campo, 
\href{http://dx.doi.org/10.1103/PhysRevA.78.045602}{Phys. Rev. A {\bf 78}, 45602 (2008)}.

\bibitem{hc-12}
Y. Hao and S. Chen, 
\href{http://dx.doi.org/10.1103/PhysRevA.86.043631}{Phys. Rev. A {\bf 86}, 43631 (2012)}.

\bibitem{Li-15}
Y. Li, 
\href{http://dx.doi.org/10.1140/epjp/i2015-15101-x}{Eur. Phys. J. Plus {\bf 130}, 101 (2015)}.

\bibitem{wrdk-14}
T. M. Wright, M. Rigol, M. J. Davis, and K. V. Kheruntsyan, 
\href{http://dx.doi.org/10.1103/PhysRevLett.113.050601}{Phys. Rev. Lett. {\bf 113}, 50601 (2014)}.

\bibitem{PiCa17} 
L. Piroli and P. Calabrese, 
\href{http://dx.doi.org/10.1103/PhysRevA.96.023611}{Phys. Rev. A {\bf 96}, 023611 (2017)}.

\bibitem{fgd-18}
F. Liu, J. R. Garrison, D.-L. Deng, Z.-.X. Gong, and A. V. Gorshkov, 
\href{https://doi.org/10.1103/PhysRevLett.121.250404}{Phys. Rev. Lett. {\bf 121}, 250404  (2018)}.



\bibitem{BlDZ08} 
I. Bloch, J. Dalibard, and W. Zwerger, 
\href{http://dx.doi.org/10.1103/RevModPhys.80.885}{Rev. Mod. Phys. {\bf 80}, 885 (2008)}.

\bibitem{GuBL13} 
X.-W. Guan, M. T. Batchelor, and C. Lee, 
\href{http://dx.doi.org/10.1103/RevModPhys.85.1633}{Rev. Mod. Phys. {\bf 85}, 1633 (2013)}.


\bibitem{Haldane1981a}
F. D. M. Haldane,
\href{https://doi.org/10.1103/PhysRevLett.47.1840}{Phys. Rev. Lett. {\bf 47}, 1840 (1981)}.

\bibitem{Haldane1981b}
F. D. M. Haldane, 
\href{https://doi.org/10.1088/0022-3719/14/19/010}{J. Phys. C {\bf14}, 2585 (1981)}.

\bibitem{Haldane1981c}
F. D. M. Haldane, 
\href{https://doi.org/10.1016/0375-9601(81)90049-9}{Phys. Lett. A {\bf 81}, 153 (1981)}.

\bibitem{Gogolin-Tsvelik}
A. O. Gogolin, A. A. Nersesyan, and A. M. Tsvelik,
\emph{Bosonization and Strongly Correlated Systems}, Cambridge (1998).

\bibitem{Giamarchi2003} T. Giamarchi, \emph{Quantum physics in one dimension}, Clarendon Press (2003).

\bibitem{Cazalilla2004}
M. A. Cazalilla,
\href{https://doi.org/10.1088/0953-4075/37/7/051}{J. Phys. B: At. Mol. Opt. Phys. {\bf 37}, S1 (2004)}


\bibitem{MaSt95} 
D. L. Maslov and M. Stone, 
\href{http://dx.doi.org/10.1103/PhysRevB.52.R5539}{Phys. Rev. B {\bf 52}, R5539 (1995)}.

\bibitem{tss}
M. Campostrini and E. Vicari, 
\href{https://doi.org/10.1103/PhysRevLett.102.240601}{Phys. Rev. Lett. {\bf 102}, 240601 (2009)};\\
M. Campostrini and E. Vicari, 
\href{https://doi.org/10.1103/PhysRevA.81.023606}{Phys. Rev. A {\bf 81}, 023606 (2010)}; \\ 
M. Campostrini and E. Vicari, 
\href{https://doi.org/10.1103/PhysRevA.81.063614}{Phys. Rev. A {\bf 81}, 063614 (2010)}.

\bibitem{Kats11} 
H. Katsura, 
\href{http://dx.doi.org/10.1088/1751-8113/44/25/252001}{J. Phys. A: Math. Theor. {\bf 44}, 252001 (2011)}.

\bibitem{HiNi11} 
T. Hikihara and T. Nishino, 
\href{http://dx.doi.org/10.1103/PhysRevB.83.060414}{Phys. Rev. B {\bf 83}, 060414 (2011)}.

\bibitem{WeRL16} 
X. Wen, S. Ryu, and A. W. W. Ludwig, 
\href{http://dx.doi.org/10.1103/PhysRevB.93.235119}{Phys. Rev. B {\bf 93}, 235119 (2016)}.

\bibitem{RDRC17} 
J. Rodr\'iguez-Laguna, J. Dubail, G. Ram\'irez, P. Calabrese, and G. Sierra, 
\href{http://dx.doi.org/10.1088/1751-8121/aa6268}{J. Phys. A: Math. Theor. {\bf 50}, 164001 (2017)}.

\bibitem{DuSC17} 
J. Dubail, J.-M. St\'ephan, and P. Calabrese, 
\href{http://dx.doi.org/10.21468/SciPostPhys.3.3.019}{SciPost Physics {\bf 3}, 019 (2017)}.

\bibitem{ToRS18} 
E. Tonni, J. Rodr\'iguez-Laguna, and G. Sierra, 
\href{http://dx.doi.org/10.1088/1742-5468/aab67d}{J. Stat. Mech. (2018) 043105}.


\bibitem{ADSV16} 
N. Allegra, J. Dubail, J.-M. Stephan, and J. Viti, 
\href{http://dx.doi.org/10.1088/1742-5468/2016/05/053108}{J. Stat. Mech.  053108 (2016)}.

\bibitem{Dubail17}
J. Dubail, J. M. St\'ephan, J. Viti, and P. Calabrese,
\href{https://doi.org/10.21468/SciPostPhys.2.1.002}{SciPost Phys. {\bf 2}, 002 (2017)}


\bibitem{Brun2017} 
Y. Brun and J. Dubail, 
\href{http://dx.doi.org/10.21468/SciPostPhys.2.2.012}{SciPost Phys. {\bf 2}, 012 (2017)}.

\bibitem{Brun2018}
Y. Brun and J. Dubail, 
\href{http://dx.doi.org/10.21468/SciPostPhys.4.6.037}{SciPost Phys. {\bf 4}, 037 (2018)}.

\bibitem{Y-thesis}
Y. Brun,  
\href{https://hal.univ-lorraine.fr/tel-02393790}{PhD thesis},\emph{Corr\`elations dans les syst\`emes quantiques inhomog\`enes \`a une dimension}, Universit\'e de Lorraine (2019).

\bibitem{j-lec}
J. Dubail, 
\href{http://theory.fi.infn.it/SFTschool/SFT_2019/LectureNotes/Dubail_lecture_0_and_1.pdf}{GGI lectures at the SFT-2019 school}, \emph{Three lectures on classical and quantum hydrodynamics 
applied to trapped 1d quantum gases}.


\bibitem{Bastianello2020}
A. Bastianello, J. Dubail, and J. M. St\`ephan,
\href{https://doi.org/10.1088/1751-8121/ab7580}{J. Phys. A: Math. Theor. {\bf 53}, 155001 (2020)}.


\bibitem{Eisler17}  
V. Eisler and D. Bauernfeind,
\href{https://doi.org/10.1103/PhysRevB.96.174301}{Phys. Rev. B {\bf 96}, 174301 (2017)}.


\bibitem{SGCI11} 
A. Shashi, L. I. Glazman, J.-S. Caux, and A. Imambekov, 
\href{http://dx.doi.org/10.1103/PhysRevB.84.045408}{Phys. Rev. B {\bf 84}, 045408 (2011)}.

\bibitem{SPCI12} 
A. Shashi, M. Panfil, J.-S. Caux, and A. Imambekov, 
\href{http://dx.doi.org/10.1103/PhysRevB.85.155136}{Phys. Rev. B {\bf 85}, 155136 (2012)}.


\bibitem{Piroli2020}
L. Piroli, S. Scopa, and P. Calabrese,
\href{https://doi.org/10.1088/1751-8121/ab94ed}{J. Phys. A: Math. Theor. 53 405001 (2020)}.

\bibitem{DiFrancesco}
P.  Di Francesco, P. Mathieu, and D. S\'en\'echal,
\emph{Conformal Field Theory}, Springer-Verlag New York (1997).

\bibitem{m-book}
G. Mussardo, {\it Statistical field theory: an introduction to exactly solved models in statistical physics}, 2nd edition, Oxford University Press (2020).

\bibitem{Vaida79}
H. G. Vaidya and C. A. Tracy,
\href{https://doi.org/10.1103/PhysRevLett.42.3}{Phys. Rev. Lett. {\bf 42}, 3 (1979)}, Erratum:
 \href{https://doi.org/10.1103/PhysRevLett.43.1540}{Phys. Rev. Lett. {\bf 43}, 1540 (1979)}

\bibitem{Vaida79b}
H. G. Vaidya and C. A. Tracy, 
\href{https://doi.org/10.1063/1.524010}{J. Math. Phys. {\bf 20}, 2291 (1979)}.

\bibitem{Bogoliubov1987}
N. M. Bogoliubov, A. G. Izergin, and N. Y.  Reshetikhin,
\href{https://doi.org/10.1088/0305-4470/20/15/047}{ J. Phys. A: Math. Gen. {\bf 20} 5361 (1987)}.

\bibitem{Colcelli18b}
A. Colcelli, G. Mussardo, and A. Trombettoni,
\href{https://doi.org/10.1209/0295-5075/122/50006}{EPL {\bf 122},  50006 (2018)};\\
A. Colcelli, J. Viti, G. Mussardo, and A. Trombettoni,
\href{https://doi.org/10.1103/PhysRevA.98.063633}{Phys. Rev. {\bf A} 98, 063633 (2018)}.


\bibitem{Ruggiero2019}
P. Ruggiero, Y. Brun, and J. Dubail,
\href{https://doi.org/10.21468/SciPostPhys.6.4.051}{SciPost Phys. {\bf 6}, 051 (2019)}

\bibitem{Maslov95}
D. L. Maslov and M. Stone,
\href{https://doi.org/10.1103/PhysRevB.52.R5539}{Phys. Rev. {\bf B} 52, R5539(R) (1995)}

\bibitem{Gangart04}
D. M. Gangardt,
\href{https://doi.org/10.1088/0305-4470/37/40/002}{J. Phys. A: Math. Gen. {\bf 37}, 9335 (2004)}.

\bibitem{col}
J. Arcila-Forero, R. Franco, and J. Silva-Valencia, 
\href{https://doi.org/10.1103/PhysRevA.94.013611}{Phys. Rev. A {\bf 94}, 013611 (2016)}.


\bibitem{Caza03} 
M. A. Cazalilla, 
\href{http://dx.doi.org/10.1103/PhysRevA.67.053606}{Phys. Rev. A {\bf 67}, 053606 (2003)}.

\bibitem{Caza04} 
M. A. Cazalilla, 
\href{http://dx.doi.org/10.1103/PhysRevA.70.041604}{Phys. Rev. A {\bf 70}, 041604 (2004)}.


\bibitem{rm-05}
M. Rigol and A. Muramatsu, 
\href{https://doi.org/10.1103/PhysRevLett.94.240403}{Phys. Rev. Lett. {\bf 94}, 240403 (2005)};\\
A. Minguzzi and D. M. Gangardt, 
\href{https://doi.org/10.1103/PhysRevLett.94.240404}{Phys. Rev. Lett. 94, 240404 (2005)}.

\bibitem{ck-12}
J.-S. Caux and R. M. Konik, 
\href{https://doi.org/10.1103/PhysRevLett.109.175301}{Phys. Rev. Lett. {\bf 109}, 175301 (2012)}.

\bibitem{csc-13} M. Collura, S. Sotiriadis, and P. Calabrese, 
\href{http://dx.doi.org/10.1103/PhysRevLett.110.245301}{Phys. Rev. Lett. {\bf 110}, 245301 (2013)};\\
M. Collura, S. Sotiriadis, and P. Calabrese, 
 \href{http://dx.doi.org/10.1088/1742-5468/2013/09/P09025}{J. Stat. Mech. (2013) P09025}.


\bibitem{RCDD20} 
P. Ruggiero, P. Calabrese, B. Doyon, and J. Dubail, 
\href{http://dx.doi.org/10.1103/PhysRevLett.124.140603}{Phys. Rev. Lett. {\bf 124}, 140603 (2020)}.

\bibitem{BCDF16} 
B. Bertini, M. Collura, J. De Nardis, and M. Fagotti, 
\href{http://dx.doi.org/10.1103/PhysRevLett.117.207201}{Phys. Rev. Lett. {\bf 117}, 207201 (2016)}.

\bibitem{CaDY16} 
O. A. Castro-Alvaredo, B. Doyon, and T. Yoshimura, 
\href{http://dx.doi.org/10.1103/PhysRevX.6.041065}{Phys. Rev. X {\bf 6}, 041065 (2016)}.





\end{thebibliography}
\end{document}